\documentclass[journal,10pt,twoside,twocolumn]{IEEEtran}

\usepackage{cite}
\usepackage[T1]{fontenc}
\usepackage{graphicx}
\usepackage{amssymb}
\usepackage{amsmath}
\usepackage{paralist}
\usepackage{setspace}
\usepackage{microtype}
\usepackage{url}
\usepackage{balance}
\usepackage[caption=false,font=footnotesize]{subfig}
\usepackage{xcolor}
\usepackage{fancyref}
\usepackage{dsfont}

\usepackage{algorithm}
\usepackage{algpseudocode}

\usepackage{amssymb}
\usepackage{amsfonts}
\usepackage{mathrsfs}
\usepackage{xspace}
\usepackage{bm}
\usepackage{upgreek}

\newcommand{\safemath}[2]{\newcommand{#1}{\ensuremath{#2}\xspace}}

\safemath{\bma}{\mathbf{a}}
\safemath{\bmb}{\mathbf{b}}
\safemath{\bmc}{\mathbf{c}}
\safemath{\bmd}{\mathbf{d}}
\safemath{\bme}{\mathbf{e}}
\safemath{\bmf}{\mathbf{f}}
\safemath{\bmg}{\mathbf{g}}
\safemath{\bmh}{\mathbf{h}}
\safemath{\bmi}{\mathbf{i}}
\safemath{\bmj}{\mathbf{j}}
\safemath{\bmk}{\mathbf{k}}
\safemath{\bml}{\mathbf{l}}
\safemath{\bmm}{\mathbf{m}}
\safemath{\bmn}{\mathbf{n}}
\safemath{\bmo}{\mathbf{o}}
\safemath{\bmp}{\mathbf{p}}
\safemath{\bmq}{\mathbf{q}}
\safemath{\bmr}{\mathbf{r}}
\safemath{\bms}{\mathbf{s}}
\safemath{\bmt}{\mathbf{t}}
\safemath{\bmu}{\mathbf{u}}
\safemath{\bmv}{\mathbf{v}}
\safemath{\bmw}{\mathbf{w}}
\safemath{\bmx}{\mathbf{x}}
\safemath{\bmy}{\mathbf{y}}
\safemath{\bmz}{\mathbf{z}}
\safemath{\bmzero}{\mathbf{0}}
\safemath{\bmone}{\mathbf{1}}

\bmdefine{\biad}{a}
\bmdefine{\bibd}{b}
\bmdefine{\bicd}{c}
\bmdefine{\bidd}{d}
\bmdefine{\bied}{e}
\bmdefine{\bifd}{f}
\bmdefine{\bigd}{g}
\bmdefine{\bihd}{h}
\bmdefine{\biid}{i}
\bmdefine{\bijd}{j}
\bmdefine{\bikd}{k}
\bmdefine{\bild}{l}
\bmdefine{\bimd}{m}
\bmdefine{\bind}{n}
\bmdefine{\biod}{o}
\bmdefine{\bipd}{p}
\bmdefine{\biqd}{q}
\bmdefine{\bird}{r}
\bmdefine{\bisd}{s}
\bmdefine{\bitd}{t}
\bmdefine{\biud}{u}
\bmdefine{\bivd}{v}
\bmdefine{\biwd}{w}
\bmdefine{\bixd}{x}
\bmdefine{\biyd}{y}
\bmdefine{\bizd}{z}

\bmdefine{\bixid}{\xi}
\bmdefine{\bilambdad}{\lambda}
\bmdefine{\bimud}{\mu}
\bmdefine{\bithetad}{\theta}
\bmdefine{\biphid}{\phi}
\bmdefine{\bideltad}{\delta}
\bmdefine{\bibetad}{\beta}

\safemath{\bmia}{\biad}
\safemath{\bmib}{\bibd}
\safemath{\bmic}{\bicd}
\safemath{\bmid}{\bidd}
\safemath{\bmie}{\bied}
\safemath{\bmif}{\bifd}
\safemath{\bmig}{\bigd}
\safemath{\bmih}{\bihd}
\safemath{\bmii}{\biid}
\safemath{\bmij}{\bijd}
\safemath{\bmik}{\bikd}
\safemath{\bmil}{\bild}
\safemath{\bmim}{\bimd}
\safemath{\bmin}{\bind}
\safemath{\bmio}{\biod}
\safemath{\bmip}{\bipd}
\safemath{\bmiq}{\biqd}
\safemath{\bmir}{\bird}
\safemath{\bmis}{\bisd}
\safemath{\bmit}{\bitd}
\safemath{\bmiu}{\biud}
\safemath{\bmiv}{\bivd}
\safemath{\bmiw}{\biwd}
\safemath{\bmix}{\bixd}
\safemath{\bmiy}{\biyd}
\safemath{\bmiz}{\bizd}

\safemath{\bmxi}{\bixid}
\safemath{\bmlambda}{\bilambdad}
\safemath{\bmmu}{\bimud}
\safemath{\bmtheta}{\bithetad}
\safemath{\bmphi}{\biphid}
\safemath{\bmdelta}{\bideltad}
\safemath{\bmbeta}{\bibetad}

\safemath{\bA}{\mathbf{A}}
\safemath{\bB}{\mathbf{B}}
\safemath{\bC}{\mathbf{C}}
\safemath{\bD}{\mathbf{D}}
\safemath{\bE}{\mathbf{E}}
\safemath{\bF}{\mathbf{F}}
\safemath{\bG}{\mathbf{G}}
\safemath{\bH}{\mathbf{H}}
\safemath{\bI}{\mathbf{I}}
\safemath{\bJ}{\mathbf{J}}
\safemath{\bK}{\mathbf{K}}
\safemath{\bL}{\mathbf{L}}
\safemath{\bM}{\mathbf{M}}
\safemath{\bN}{\mathbf{N}}
\safemath{\bO}{\mathbf{O}}
\safemath{\bP}{\mathbf{P}}
\safemath{\bQ}{\mathbf{Q}}
\safemath{\bR}{\mathbf{R}}
\safemath{\bS}{\mathbf{S}}
\safemath{\bT}{\mathbf{T}}
\safemath{\bU}{\mathbf{U}}
\safemath{\bV}{\mathbf{V}}
\safemath{\bW}{\mathbf{W}}
\safemath{\bX}{\mathbf{X}}
\safemath{\bY}{\mathbf{Y}}
\safemath{\bZ}{\mathbf{Z}}

\safemath{\bZero}{\mathbf{0}}
\safemath{\bOne}{\mathbf{1}}
\safemath{\bDelta}{\mathbf{\Delta}}
\safemath{\bLambda}{\mathbf{\UpLambda}}
\safemath{\bPhi}{\mathbf{\Upphi}}
\safemath{\bSigma}{\mathbf{\Upsigma}}
\safemath{\bOmega}{\mathbf{\Upomega}}
\safemath{\bTheta}{\mathbf{\Uptheta}}

\bmdefine{\biAd}{A}
\bmdefine{\biBd}{B}
\bmdefine{\biCd}{C}
\bmdefine{\biDd}{D}
\bmdefine{\biEd}{E}
\bmdefine{\biFd}{F}
\bmdefine{\biGd}{G}
\bmdefine{\biHd}{H}
\bmdefine{\biId}{I}
\bmdefine{\biJd}{J}
\bmdefine{\biKd}{K}
\bmdefine{\biLd}{L}
\bmdefine{\biMd}{M}
\bmdefine{\biOd}{N}
\bmdefine{\biPd}{O}
\bmdefine{\biQd}{P}
\bmdefine{\biRd}{R}
\bmdefine{\biSd}{S}
\bmdefine{\biTd}{T}
\bmdefine{\biUd}{U}
\bmdefine{\biVd}{V}
\bmdefine{\biWd}{W}
\bmdefine{\biXd}{X}
\bmdefine{\biYd}{Y}
\bmdefine{\biZd}{Z}

\bmdefine{\biDelta}{\Delta}
\bmdefine{\biLambda}{\Lambda}
\bmdefine{\biPhi}{\Phi}
\bmdefine{\biSigma}{\Sigma}
\bmdefine{\biOmega}{\Omega}
\bmdefine{\biTheta}{\Theta}

\safemath{\bimA}{\biAd}
\safemath{\bimB}{\biBd}
\safemath{\bimC}{\biCd}
\safemath{\bimD}{\biDd}
\safemath{\bimE}{\biEd}
\safemath{\bimF}{\biFd}
\safemath{\bimG}{\biGd}
\safemath{\bimH}{\biHd}
\safemath{\bimI}{\biId}
\safemath{\bimJ}{\biJd}
\safemath{\bimK}{\biKd}
\safemath{\bimL}{\biLd}
\safemath{\bimM}{\biMd}
\safemath{\bimN}{\biNd}
\safemath{\bimO}{\biOd}
\safemath{\bimP}{\biPd}
\safemath{\bimQ}{\biQd}
\safemath{\bimR}{\biRd}
\safemath{\bimS}{\biSd}
\safemath{\bimT}{\biTd}
\safemath{\bimU}{\biUd}
\safemath{\bimV}{\biVd}
\safemath{\bimW}{\biWd}
\safemath{\bimX}{\biXd}
\safemath{\bimY}{\biYd}
\safemath{\bimZ}{\biZd}

\safemath{\bimDelta}{\biDelta}
\safemath{\bimLambda}{\biLambda}
\safemath{\bimPhi}{\biPhi}
\safemath{\bimSigma}{\biSigma}
\safemath{\bimOmega}{\biOmega}
\safemath{\bimTheta}{\biTheta}

\safemath{\setA}{\mathcal{A}}
\safemath{\setB}{\mathcal{B}}
\safemath{\setC}{\mathcal{C}}
\safemath{\setD}{\mathcal{D}}
\safemath{\setE}{\mathcal{E}}
\safemath{\setF}{\mathcal{F}}
\safemath{\setG}{\mathcal{G}}
\safemath{\setH}{\mathcal{H}}
\safemath{\setI}{\mathcal{I}}
\safemath{\setJ}{\mathcal{J}}
\safemath{\setK}{\mathcal{K}}
\safemath{\setL}{\mathcal{L}}
\safemath{\setM}{\mathcal{M}}
\safemath{\setN}{\mathcal{N}}
\safemath{\setO}{\mathcal{O}}
\safemath{\setP}{\mathcal{P}}
\safemath{\setQ}{\mathcal{Q}}
\safemath{\setR}{\mathcal{R}}
\safemath{\setS}{\mathcal{S}}
\safemath{\setT}{\mathcal{T}}
\safemath{\setU}{\mathcal{U}}
\safemath{\setV}{\mathcal{V}}
\safemath{\setW}{\mathcal{W}}
\safemath{\setX}{\mathcal{X}}
\safemath{\setY}{\mathcal{Y}}
\safemath{\setZ}{\mathcal{Z}}
\safemath{\emptySet}{\varnothing}

\safemath{\colA}{\mathscr{A}}
\safemath{\colB}{\mathscr{B}}
\safemath{\colC}{\mathscr{C}}
\safemath{\colD}{\mathscr{D}}
\safemath{\colE}{\mathscr{E}}
\safemath{\colF}{\mathscr{F}}
\safemath{\colG}{\mathscr{G}}
\safemath{\colH}{\mathscr{H}}
\safemath{\colI}{\mathscr{I}}
\safemath{\colJ}{\mathscr{J}}
\safemath{\colK}{\mathscr{K}}
\safemath{\colL}{\mathscr{L}}
\safemath{\colM}{\mathscr{M}}
\safemath{\colN}{\mathscr{N}}
\safemath{\colO}{\mathscr{O}}
\safemath{\colP}{\mathscr{P}}
\safemath{\colQ}{\mathscr{Q}}
\safemath{\colR}{\mathscr{R}}
\safemath{\colS}{\mathscr{S}}
\safemath{\colT}{\mathscr{T}}
\safemath{\colU}{\mathscr{U}}
\safemath{\colV}{\mathscr{V}}
\safemath{\colW}{\mathscr{W}}
\safemath{\colX}{\mathscr{X}}
\safemath{\colY}{\mathscr{Y}}
\safemath{\colZ}{\mathscr{Z}}

\safemath{\opA}{\mathbb{A}}
\safemath{\opB}{\mathbb{B}}
\safemath{\opC}{\mathbb{C}}
\safemath{\opD}{\mathbb{D}}
\safemath{\opE}{\mathbb{E}}
\safemath{\opF}{\mathbb{F}}
\safemath{\opG}{\mathbb{G}}
\safemath{\opH}{\mathbb{H}}
\safemath{\opI}{\mathbb{I}}
\safemath{\opJ}{\mathbb{J}}
\safemath{\opK}{\mathbb{K}}
\safemath{\opL}{\mathbb{L}}
\safemath{\opM}{\mathbb{M}}
\safemath{\opN}{\mathbb{N}}
\safemath{\opO}{\mathbb{O}}
\safemath{\opP}{\mathbb{P}}
\safemath{\opQ}{\mathbb{Q}}
\safemath{\opR}{\mathbb{R}}
\safemath{\opS}{\mathbb{S}}
\safemath{\opT}{\mathbb{T}}
\safemath{\opU}{\mathbb{U}}
\safemath{\opV}{\mathbb{V}}
\safemath{\opW}{\mathbb{W}}
\safemath{\opX}{\mathbb{X}}
\safemath{\opY}{\mathbb{Y}}
\safemath{\opZ}{\mathbb{Z}}
\safemath{\opZero}{\mathbb{O}}
\safemath{\identityop}{\opI}

\safemath{\veca}{\bma}
\safemath{\vecb}{\bmb}
\safemath{\vecc}{\bmc}
\safemath{\vecd}{\bmd}
\safemath{\vece}{\bme}
\safemath{\vecf}{\bmf}
\safemath{\vecg}{\bmg}
\safemath{\vech}{\bmh}
\safemath{\veci}{\bmi}
\safemath{\vecj}{\bmj}
\safemath{\veck}{\bmk}
\safemath{\vecl}{\bml}
\safemath{\vecm}{\bmm}
\safemath{\vecn}{\bmn}
\safemath{\veco}{\bmo}
\safemath{\vecp}{\bmp}
\safemath{\vecq}{\bmq}
\safemath{\vecr}{\bmr}
\safemath{\vecs}{\bms}
\safemath{\vect}{\bmt}
\safemath{\vecu}{\bmu}
\safemath{\vecv}{\bmv}
\safemath{\vecw}{\bmw}
\safemath{\vecx}{\bmx}
\safemath{\vecy}{\bmy}
\safemath{\vecz}{\bmz}

\safemath{\veczero}{\bmzero}
\safemath{\vecone}{\bmone}
\safemath{\vecxi}{\bmxi}
\safemath{\veclambda}{\bmlambda}
\safemath{\vecmu}{\bmmu}
\safemath{\vectheta}{\bmtheta}
\safemath{\vecphi}{\bmphi}
\safemath{\vecdelta}{\bmdelta}
\safemath{\vecbeta}{\bmbeta}

\safemath{\matA}{\bA}
\safemath{\matB}{\bB}
\safemath{\matC}{\bC}
\safemath{\matD}{\bD}
\safemath{\matE}{\bE}
\safemath{\matF}{\bF}
\safemath{\matG}{\bG}
\safemath{\matH}{\bH}
\safemath{\matI}{\bI}
\safemath{\matJ}{\bJ}
\safemath{\matK}{\bK}
\safemath{\matL}{\bL}
\safemath{\matM}{\bM}
\safemath{\matN}{\bN}
\safemath{\matO}{\bO}
\safemath{\matP}{\bP}
\safemath{\matQ}{\bQ}
\safemath{\matR}{\bR}
\safemath{\matS}{\bS}
\safemath{\matT}{\bT}
\safemath{\matU}{\bU}
\safemath{\matV}{\bV}
\safemath{\matW}{\bW}
\safemath{\matX}{\bX}
\safemath{\matY}{\bY}
\safemath{\matZ}{\bZ}
\safemath{\matzero}{\bmzero}

\safemath{\matDelta}{\bDelta}
\safemath{\matLambda}{\bLambda}
\safemath{\matPhi}{\bPhi}
\safemath{\matSigma}{\bSigma}
\safemath{\matOmega}{\bOmega}
\safemath{\matTheta}{\bTheta}

\safemath{\matidentity}{\matI}
\safemath{\matone}{\matO}

\safemath{\rnda}{A}
\safemath{\rndb}{B}
\safemath{\rndc}{C}
\safemath{\rndd}{D}
\safemath{\rnde}{E}
\safemath{\rndf}{F}
\safemath{\rndg}{G}
\safemath{\rndh}{H}
\safemath{\rndi}{I}
\safemath{\rndj}{J}
\safemath{\rndk}{K}
\safemath{\rndl}{L}
\safemath{\rndm}{M}
\safemath{\rndn}{N}
\safemath{\rndo}{O}
\safemath{\rndp}{P}
\safemath{\rndq}{Q}
\safemath{\rndr}{R}
\safemath{\rnds}{S}
\safemath{\rndt}{T}
\safemath{\rndu}{U}
\safemath{\rndv}{V}
\safemath{\rndw}{W}
\safemath{\rndx}{X}
\safemath{\rndy}{Y}
\safemath{\rndz}{Z}

\safemath{\rveca}{\bimA}
\safemath{\rvecb}{\bimB}
\safemath{\rvecc}{\bimC}
\safemath{\rvecd}{\bimD}
\safemath{\rvece}{\bimE}
\safemath{\rvecf}{\bimF}
\safemath{\rvecg}{\bimG}
\safemath{\rvech}{\bimH}
\safemath{\rveci}{\bimI}
\safemath{\rvecj}{\bimJ}
\safemath{\rveck}{\bimK}
\safemath{\rvecl}{\bimL}
\safemath{\rvecm}{\bimM}
\safemath{\rvecn}{\bimN}
\safemath{\rveco}{\bomO}
\safemath{\rvecp}{\bimP}
\safemath{\rvecq}{\bimQ}
\safemath{\rvecr}{\bimR}
\safemath{\rvecs}{\bimS}
\safemath{\rvect}{\bimT}
\safemath{\rvecu}{\bimU}
\safemath{\rvecv}{\bimV}
\safemath{\rvecw}{\bimW}
\safemath{\rvecx}{\bimX}
\safemath{\rvecy}{\bimY}
\safemath{\rvecz}{\bimZ}

\safemath{\rvecxi}{\bmxi}
\safemath{\rveclambda}{\bmlambda}
\safemath{\rvecmu}{\bmmu}
\safemath{\rvectheta}{\bmtheta}
\safemath{\rvecphi}{\bmphi}

\safemath{\rmatA}{\bimA}
\safemath{\rmatB}{\bimB}
\safemath{\rmatC}{\bimC}
\safemath{\rmatD}{\bimD}
\safemath{\rmatE}{\bimE}
\safemath{\rmatF}{\bimF}
\safemath{\rmatG}{\bimG}
\safemath{\rmatH}{\bimH}
\safemath{\rmatI}{\bimI}
\safemath{\rmatJ}{\bimJ}
\safemath{\rmatK}{\bimK}
\safemath{\rmatL}{\bimL}
\safemath{\rmatM}{\bimM}
\safemath{\rmatN}{\bimN}
\safemath{\rmatO}{\bimO}
\safemath{\rmatP}{\bimP}
\safemath{\rmatQ}{\bimQ}
\safemath{\rmatR}{\bimR}
\safemath{\rmatS}{\bimS}
\safemath{\rmatT}{\bimT}
\safemath{\rmatU}{\bimU}
\safemath{\rmatV}{\bimV}
\safemath{\rmatW}{\bimW}
\safemath{\rmatX}{\bimX}
\safemath{\rmatY}{\bimY}
\safemath{\rmatZ}{\bimZ}

\safemath{\rmatDelta}{\bimDelta}
\safemath{\rmatLambda}{\bimLambda}
\safemath{\rmatPhi}{\bimPhi}
\safemath{\rmatSigma}{\bimSigma}
\safemath{\rmatOmega}{\bimOmega}
\safemath{\rmatTheta}{\bimTheta}

\usepackage{amssymb}
\usepackage{amsfonts}
\usepackage{mathrsfs}
\usepackage{xspace}
\usepackage{bm}
\usepackage{fancyref}
\usepackage{textcomp}

\usepackage{multirow}
\usepackage{stmaryrd}

\newenvironment{textbmatrix}{	\setlength{\arraycolsep}{2.5pt}%
								\big[\begin{matrix}}{\end{matrix}\big]%
								\raisebox{0.08ex}{\vphantom{M}}}

\def\be{\begin{equation}}
\def\ee{\end{equation}}
\def\een{\nonumber \end{equation}}
\def\mat{\begin{bmatrix}}
\def\emat{\end{bmatrix}}
\def\btm{\begin{textbmatrix}}
\def\etm{\end{textbmatrix}}

\def\ba#1\ea{\begin{align}#1\end{align}}
\def\bas#1\eas{\begin{align*}#1\end{align*}}
\def\bs#1\es{\begin{split}#1\end{split}} 
\def\bg#1\eg{\begin{gather}#1\end{gather}}
\def\bml#1\eml{\begin{multline}#1\end{multline}}
\def\bi#1\ei{\begin{itemize}#1\end{itemize}}

\newcommand{\lefto}{\mathopen{}\left}

\DeclareMathOperator{\tr}{tr}				
\DeclareMathOperator{\rank}{rank}			
\DeclareMathOperator{\sign}{sgn}			
\DeclareMathOperator*{\argmin}{arg\;min}		
\DeclareMathOperator{\Exop}{\opE}			

\newcommand{\Ex}[2]{\ensuremath{\Exop_{#1}\lefto[#2\right]}} 	
\newcommand{\abs}[1]{\lefto\lvert#1\right\rvert}		

\newcommand{\vecnorm}[1]{\lefto\lVert#1\right\rVert}		

\safemath{\dirac}{\delta}					
\safemath{\krond}{\dirac}					

\safemath{\upto}{\uparrow}
\safemath{\downto}{\downarrow}
\safemath{\iu}{j}							
\safemath{\ev}{\lambda}						
\safemath{\hilseqspace}{l^{2}}				
\newcommand{\banachfunspace}[1]{\setL^{#1}}	
\safemath{\hilfunspace}{\banachfunspace{2}}	

\safemath{\SNR}{\textsf{SNR}} 				
\safemath{\PAR}{\textsf{PAR}} 				
\safemath{\No}{N_0}							
\safemath{\Es}{E_s}							
\safemath{\Eb}{E_b}							
\safemath{\EbNo}{\frac{\Eb}{\No}}
\safemath{\EsNo}{\frac{\Es}{\No}}

\DeclareMathOperator{\CHop}{\ensuremath{\opH}} 
\safemath{\tvir}{\rndh_{\CHop}}				
\safemath{\tvtf}{\rndl_{\CHop}}				
\safemath{\spf}{\rnds_{\CHop}}				
\safemath{\bff}{H_{\CHop}}					

\safemath{\ircf}{r_{h}}						
\safemath{\tftvcf}{r_{s}}					
\safemath{\tfcf}{r_{l}}						
\safemath{\bfcf}{r_{H}}						

\safemath{\tcorr}{c_h}						
\safemath{\scf}{c_{s}}						
\safemath{\tfcorr}{c_{l}}					
\safemath{\fcorr}{c_{H}}						

\safemath{\mi}{I}							
\safemath{\capacity}{C}						

\safemath{\normal}{\mathcal{N}}			
\safemath{\jpg}{\mathcal{CN}}			
\safemath{\mchain}{\leftrightarrow}		
\newcommand{\given}{\,\vert\,}				

\safemath{\dB}{\,\mathrm{dB}}
\safemath{\dBm}{\,\mathrm{dBm}}
\safemath{\Hz}{\,\mathrm{Hz}}
\safemath{\kHz}{\,\mathrm{kHz}}
\safemath{\MHz}{\,\mathrm{MHz}}
\safemath{\GHz}{\,\mathrm{GHz}}
\safemath{\s}{\,\mathrm{s}}
\safemath{\ms}{\,\mathrm{ms}}
\safemath{\mus}{\,\mathrm{\text{\textmu}s}}
\safemath{\ns}{\,\mathrm{ns}}
\safemath{\ps}{\,\mathrm{ps}}
\safemath{\meter}{\,\mathrm{m}}
\safemath{\mm}{\,\mathrm{mm}}
\safemath{\cm}{\,\mathrm{cm}}
\safemath{\m}{\,\mathrm{m}}
\safemath{\W}{\,\mathrm{W}}
\safemath{\mW}{\, \mathrm{mW}}
\safemath{\J}{\,\mathrm{J}}
\safemath{\K}{\,\mathrm{K}}
\safemath{\bit}{\,\mathrm{bit}}
\safemath{\nat}{\,\mathrm{nat}}

\safemath{\define}{\triangleq}			

\safemath{\equivalent}{\sim}
\safemath{\distas}{\sim}					
\safemath{\sdiff}{\Delta}				

\safemath{\reals}{\mathbb{R}}
\safemath{\positivereals}{\reals_{+}}
\safemath{\integers}{\mathbb{Z}}
\safemath{\posint}{\integers_{+}}
\safemath{\naturals}{\mathbb{N}}
\safemath{\posnaturals}{\naturals_{+}}
\safemath{\complexset}{\mathbb{C}}
\safemath{\rationals}{\mathbb{Q}}

\newcommand*{\fancyrefapplabelprefix}{app}		
\newcommand*{\fancyrefthmlabelprefix}{thm}		
\newcommand*{\fancyreflemlabelprefix}{lem}		
\newcommand*{\fancyrefcorlabelprefix}{cor}		
\newcommand*{\fancyrefdeflabelprefix}{def}		
\newcommand*{\fancyrefproplabelprefix}{prop}	
\newcommand*{\fancyrefobslabelprefix}{obs}		
\newcommand*{\fancyrefalglabelprefix}{alg}		
\newcommand*{\fancyrefasmlabelprefix}{asm}	    

\frefformat{vario}{\fancyrefseclabelprefix}{Section~#1}
\frefformat{vario}{\fancyrefthmlabelprefix}{Theorem~#1}
\frefformat{vario}{\fancyreflemlabelprefix}{Lemma~#1}
\frefformat{vario}{\fancyrefcorlabelprefix}{Corollary~#1}
\frefformat{vario}{\fancyrefdeflabelprefix}{Definition~#1}
\frefformat{vario}{\fancyrefobslabelprefix}{Observation~#1}
\frefformat{vario}{\fancyrefasmlabelprefix}{Assumption~#1}
\frefformat{vario}{\fancyreffiglabelprefix}{Fig.~#1}
\frefformat{vario}{\fancyrefapplabelprefix}{Appendix~#1} 
\frefformat{vario}{\fancyrefproplabelprefix}{Proposition~#1}
\frefformat{vario}{\fancyrefalglabelprefix}{Algorithm~#1}
\frefformat{vario}{\fancyrefeqlabelprefix}{(#1)}

\newtheorem{thm}{Theorem}
\newtheorem{cor}[thm]{Corollary}   

\safemath{\dictab}{[\,\dicta\,\,\dictb\,]}

\safemath{\ysig}{\bmy}
\safemath{\ysighat}{\hat{\ysig}}
\safemath{\ysigdim}{M}
\safemath{\xsig}{\bmx}
\safemath{\xsigdim}{N}
\safemath{\nx}{n_x}
\safemath{\zsig}{\bmz}
\safemath{\zsigdim}{\ysigdim}
\safemath{\rsig}{\bmr}
\safemath{\Adict}{\bA}
\safemath{\Adicttilde}{\widetilde{\Adict}}
\safemath{\Adictdim}{\outputdim\times\xsigdim}
\safemath{\avec}{\bma}
\safemath{\avectilde}{\tilde{\avec}}
\safemath{\Bdict}{\bB}
\safemath{\Bdicttilde}{\widetilde{\Bdict}}
\safemath{\Cdict}{\bC}
\safemath{\cvec}{\bmc}
\safemath{\Ddict}{\bD}
\safemath{\Ddictdim}{\ysigdim\times\xsigdim}
\safemath{\dvec}{\bmd}
\safemath{\Ddicttilde}{\widetilde{\bD}}
\safemath{\Bonb}{\bB}
\safemath{\bvec}{\bmb}
\safemath{\Bonbdim}{\ysigdim\times\ysigdim}
\safemath{\noise}{\bmn}
\safemath{\noisedim}{\ysigim}
\safemath{\err}{\bme}
\safemath{\errdim}{\ysigdim}
\safemath{\errset}{\setE}
\safemath{\nerr}{n_e}
\safemath{\delop}{\bP_\errset}
\safemath{\delopc}{\bP_{{\errset}^c}}

\safemath{\cplxi}{\imath}
\safemath{\cplxj}{\jmath}

\safemath{\dict}{\matD}
\safemath{\inputdim}{N}		
\safemath{\outputdim}{M}		
\safemath{\sparsity}{S}	
\safemath{\inputdimA}{{N_a}}	
\safemath{\inputdimB}{{N_b}}	
\safemath{\elemA}{{n_a}}	
\safemath{\elemB}{{n_b}}	
\safemath{\resA}{\matR_a}	
\safemath{\resB}{\matR_b}	
\safemath{\subD}{\matS} 
\safemath{\subA}{\matS_a} 
\safemath{\subB}{\matS_b} 
\safemath{\dicta}{\matA} 	
\safemath{\dictb}{\matB} 	
\safemath{\hollowS}{H}
\safemath{\hollowA}{H_a}
\safemath{\hollowB}{H_b}
\safemath{\cross}{Z}
\safemath{\coh}{\mu_d}			
\safemath{\coha}{\mu_a}			
\safemath{\cohb}{\mu_b}			
\safemath{\mubs}{\nu}	
\safemath{\cohm}{\mu_m} 
\safemath{\dictset}{\setD}	
\safemath{\dictsetp}{\dictset(\coh,\coha,\cohb)}	
\safemath{\dictsetgen}{\dictset_\text{gen}}
\safemath{\dictsetgenp}{\dictsetgen(\coh)}
\safemath{\dictsetonb}{\dictset_\text{onb}}
\safemath{\dictsetonbp}{\dictsetonb(\coh)}

\safemath{\leftside}{U}
\safemath{\rightsideA}{R_a}
\safemath{\rightsideB}{R_b}

\safemath{\indexS}{\setI_S} 

\safemath{\na}{n_a}			
\safemath{\nb}{n_b}			
\safemath{\coeffa}{p_i}	
\safemath{\coeffb}{q_j}	
\safemath{\seta}{\setP}		
\safemath{\setb}{\setQ}     
\safemath{\setw}{\setW}	
\safemath{\setz}{\setZ}	
\safemath{\cola}{\veca}		
\safemath{\colb}{\vecb}		
\safemath{\cold}{\vecd}		
\safemath{\inputvec}{\vecx} 	
\safemath{\error}{\vece}	
\safemath{\noiseout}{\vecz} 	
\safemath{\inputvecel}{x}
\safemath{\inputveca}{\vecx_a}
\safemath{\inputvecb}{\vecx_b}
\safemath{\outputvec}{\vecy}	
\safemath{\lambdamin}{\lambda_{\mathrm{min}}}

\safemath{\elltwo}{\ell_2}
\safemath{\ellone}{\ell_1}
\safemath{\ellzero}{\ell_0}
\safemath{\ellinf}{\ell_\infty}
\safemath{\ellinftilde}{\ell_{\widetilde\infty}}
\safemath{\licard}{Z(\coh,\coha,\cohb)}
\safemath{\xsol}{\hat{x}}
\safemath{\xbord}{x_b}		
\safemath{\xstat}{x_s}		
\safemath{\xstatLone}{\tilde{x}_s}
\safemath{\order}{\mathcal{O}} 
\safemath{\scales}{\Theta} 
\safemath{\ones}{\mathbf{1}} 
\safemath{\zeroes}{\mathbf{0}} 
\safemath{\thlone}{\kappa(\coh,\cohb)} 
\safemath{\constoneA}{\delta} 
\safemath{\constoneB}{\epsilon} 
\safemath{\nlarge}{L}				   
\safemath{\sumlarge}{S_\nlarge}
\safemath{\maxlarger}{P_\nlarge}	   
\safemath{\Pzero}{\textrm{P0}}	
\safemath{\Pone}{\textrm{P1}}
\safemath{\vecfir}{\vecw}			 
\safemath{\vecsec}{\vecz}
\safemath{\elvecfir}{w}              
\safemath{\elvecsec}{z}				 
\safemath{\nlargefir}{n}
\safemath{\normout}{\gamma}
\safemath{\auxfun}{h}
\safemath{\supp}{\textrm{supp}}

\safemath{\indexa}{\ell}
\safemath{\indexb}{r}
\safemath{\indexc}{i}
\safemath{\indexd}{j}

\safemath{\project}{P}

\newcommand{\LQP}{{\text{LQP}}}
\newcommand{\OBQP}{{\text{QP}}}
\newcommand{\SDR}{{\text{SDR-QP}}}
\newcommand{\LINF}{{\ell_\infty^2\!\text{-QP}}}
\newcommand{\SP}{{\text{SP}}}

\newcommand{\quantize}{\mathcal{Q}}
\newcommand{\precode}{\mathcal{P}}

\DeclareMathOperator{\diag}{diag}

\newcommand{\snr}{\rho} 
\newcommand{\sindr}{\gamma}

\IEEEoverridecommandlockouts 
\allowdisplaybreaks 
\usepackage{color}

\usepackage{color}
\definecolor{links}{rgb}{0.7,0,0}   
\definecolor{urls}{rgb}{0,0,0.8}    
\definecolor{cites}{rgb}{0,0,0.8}   
\usepackage[colorlinks,hyperindex,linkcolor=black,citecolor=black,urlcolor=black]{hyperref} 

\begin{document}

\title{Quantized Precoding for Massive MU-MIMO}
\author{\normalsize Sven Jacobsson,~\IEEEmembership{\normalsize Student Member,~IEEE},\thanks{S.\ Jacobsson is with Ericsson Research and Chalmers University of Technology, Gothenburg, Sweden (e-mail: \url{sven.jacobsson@ericsson.com}).} Giuseppe Durisi,~\IEEEmembership{\normalsize Senior Member,~IEEE},\thanks{G.\ Durisi is with Chalmers University of Technology, Gothenburg, Sweden (e-mail: \url{durisi@chalmers.se}).}  Mikael Coldrey,~\IEEEmembership{\normalsize Member,~IEEE},\thanks{M.\ Coldrey is with Ericsson Research, Gothenburg, Sweden (e-mail: \url{mikael.coldrey@ericsson.com})} Tom Goldstein,~\IEEEmembership{\normalsize Member,~IEEE},\thanks{T. Goldstein is with the Department of Computer Science, University of Maryland, College Park, MD (e-mail: \url{ tomg@cs.umd.edu}).} and Christoph Studer,~\IEEEmembership{\normalsize Senior Member,~IEEE}\thanks{C.~Studer is with the School of Electrical and Computer Engineering, Cornell University, Ithaca, NY (e-mail: \url{studer@cornell.edu}).}\thanks{The work of S.~Jacobsson and G.~Durisi was supported by the Swedish Foundation for Strategic Research under grant ID14-0022, and by the Swedish Governmental Agency for Innovation Systems (VINNOVA) within the competence center~ChaseOn.}\thanks{The work of T.~Goldstein was supported in part by the US National Science Foundation (NSF) under grant CCF-1535902 and by the US Office of Naval Research under grant N00014-17-1-2078.}\thanks{The work of C.~Studer was supported in part by Xilinx Inc.\ and by the US  NSF under grants ECCS-1408006, CCF-1535897, and CAREER CCF-1652065.}\thanks{The system simulator for the precoders studied in this paper is available on GitHub: \url{https://github.com/quantizedmassivemimo/1bit_precoding}.}}
\maketitle


\begin{abstract}
Massive multiuser (MU) multiple-input multiple-output (MIMO) is foreseen to be one of the key technologies in fifth-generation wireless communication systems.
In this paper, we investigate the problem of downlink precoding for a narrowband massive MU-MIMO system with low-resolution digital-to-analog converters (DACs) at the base station (BS). We analyze the performance of linear precoders, such as maximal-ratio transmission and zero-forcing, subject to coarse quantization. Using Bussgang's theorem, we derive a closed-form approximation on the rate achievable under such coarse quantization. 
Our results reveal that the performance attainable with infinite-resolution DACs can be approached using DACs having only~$3$~to~$4$\,bits of resolution, depending on the number of BS antennas and the number of user equipments (UEs).
For the case of 1-bit DACs,  we also propose novel nonlinear precoding algorithms that significantly outperform  linear precoders at the cost of an increased computational complexity. 
Specifically, we show that nonlinear precoding incurs only a $3$\,dB penalty compared to the infinite-resolution case for an uncoded bit error rate of $10^{-3}$, in a system with $128$~BS antennas that uses 1-bit DACs and serves $16$ single-antenna UEs. In contrast, the penalty for linear precoders is about $8$\,dB.
\end{abstract}




\section{Introduction}
\label{sec:introduction}

Massive multiuser (MU) multiple-input multiple-output (MIMO) wireless systems, where the base station (BS) is equipped with several hundreds of antenna elements, promises significant improvements in spectral efficiency, energy efficiency, reliability, and coverage compared to traditional cellular systems~\cite{rusek14a, larsson14a, lu14a}. 
Increasing the number of radio frequency~(RF) chains at the BS could, however, lead to significant increases in hardware complexity, system costs, and circuit power consumption. Therefore, practical massive MU-MIMO systems may require low-cost and power-efficient hardware components at the~BS.
%
%
In this paper, we consider the downlink of massive MU-MIMO system, where the BS is equipped with low-resolution digital-to-analog converters (DACs) and transmits data to multiple, independent user equipments (UEs) in the same time-frequency resource.

For the quantization-free case (infinite-resolution DACs), the capacity region of the MU downlink Gaussian channel has been characterized in \cite{caire03a, yu04a, viswanath03a, viswanath03b}. When channel state information (CSI) is known noncausally at the BS, dirty-paper coding~(DPC)~\cite{costa83a} is known to achieve the sum-rate capacity~\cite{viswanath03a}. 
Several precoding algorithms to approach the DPC performance have been proposed (see, e.g., \cite{wesel98a, windpassinger04a, windpassinger04b, peel05b}).
Most of these precoding methods are, however, computationally demanding, and their complexity scales unfavorably with the number of BS~antennas, preventing their use in massive MU-MIMO.
Linear precoding, on the other hand, is an attractive low-complexity approach to massive MU-MIMO downlink precoding, which offers competitive performance to DPC for large antenna arrays~\cite{yang13c,ngo13b}. 

These results assume that the RF circuitry connected to each antenna port at the BS is ideal. The impact of RF hardware impairments at the transmit side has been investigated in, e.g.,~\cite{Studer_Tx_OFDM,gustavsson14a, zhang15d, athley15a}. Some of these results indicate that massive MU-MIMO exhibits a certain degree of resilience against RF impairments. The crude aggregate models used for characterizing such  hardware impairments, however, are unable to accurately capture the distortion caused by low-resolution DACs. 

\subsection{What are the Benefits of Quantized Massive MU-MIMO?}

One of the dominant sources of power consumption in massive MU-MIMO systems are the data converters at the~BS. In the downlink, the transmit baseband signal at each RF chain is generated by a pair of DACs.
The power consumption of these DACs increases exponentially with the resolution (in bits) and linearly with the bandwidth~\cite{walden99a, murmanna}.
In traditional multi-antenna systems, each RF port is connected to a pair of high-resolution DACs (e.g., 10-bit or more). For massive MU-MIMO systems with hundreds or even thousands of  antenna elements, this would lead to prohibitively high power consumption due to the large number of required DACs. Hence, the DAC resolution must be limited to keep the power budget within tolerable levels.
%
%
Furthermore, an often overlooked issue in massive MU-MIMO is the vast amount of data that must be exchanged between the baseband-processing unit and the radio unit (where the DACs are located). To make matters worse, in many deployment scenarios, these two units are separated by a large distance. Hence, lowering the DAC resolution is a potential solution to mitigate the data-rate bottleneck on the fronthaul.

\subsection{Relevant Prior Art}

\subsubsection{Quantized Receivers} Reducing the fronthaul throughput at the BS can be achieved by using low-resolution DACs in the downlink and low-resolution analog-to-digital converters~(ADCs) in the uplink.
Several recent contributions have studied the use of low-resolution ADCs in the massive MU-MIMO {uplink}. In particular, there has been a significant interest in the 1-bit ADC case.
For frequency-flat channels, the performance of 1-bit ADCs followed by linear detectors was analyzed in, e.g., \cite{risi14a, jacobsson15a, jacobsson17b, li17b}, where it was shown that large achievable sum rates are supported.
Similar conclusions were made in \cite{mollen16c} for the frequency-selective case. 
Nonlinear detection algorithms for frequency-selective channels were studied in, e.g.,~\cite{studer16a}. These results suggest that the number of ADC bits can be reduced significantly compared to today's~systems.


\subsubsection{Quantized Precoding}
In contrast to the uplink case, there has only been a small number of contributions that consider the massive MU-MIMO {downlink} with low-resolution DACs at the BS.
In~\cite{mezghani09c}, the authors design a linear-quantized precoder based on the minimum mean-square error (MMSE) criterion, taking into account the distortion caused by the DACs. For DACs with $4$ to $6$ bits resolution, the precoder proposed in~\cite{mezghani09c} is shown to outperform conventional linear-quantized precoders for small-to-moderate-sized MIMO systems at high signal-to-noise ratio (SNR). 
Massive MU-MIMO systems with 1-bit DACs are investigated in \cite{guerreiro16a}, where it is shown that maximal ratio transmission (MRT) precoding results in manageable distortion levels.
Again for the case of $1$-bit DACs, the authors of \cite{saxena16a} analyze the performance of zero-forcing (ZF) precoding on a Rayleigh-fading channel. Interestingly, it is shown that the received signal can be made proportional to the transmitted signal when the number of BS antennas tend to infinity. This implies that the severe per-antenna distortion caused by the $1$-bit DACs averages out when many transmit antennas are available. 
A linear precoder where the $1$-bit quantized outcomes are rescaled in the analog domain was presented in~\cite{usman16a}. There, the authors use the gradient projection algorithm to find a precoder that yields improved performance over the one reported~in~\cite{mezghani09c}. 
In \cite{zhang16a}, it is shown that, in the presence of transceiver nonlinearities (e.g., finite-resolution DACs), the achievable rate can be improved by minimizing the MSE between the transmitted symbols and the received signal prior to decoding. This result which, as we shall see, is related to the approach taken in this paper, relies on the assumptions of Gaussian inputs and nearest-neighbor decoding.

\subsubsection{Low-PAR and Constant-Envelope Precoding} Other types of hardware-aware precoding have previously been considered for massive MU-MIMO systems, with the goal of reducing the linearity requirements at the BS.
In~\cite{studer13a}, joint MU precoding and peak-to-average power ratio~(PAR) reduction was achieved by solving a convex optimization problem. Constant-envelope precoding, which minimizes the PAR by transmitting constant-modulus signals only, was studied in~\cite{mohammed13a,mohammed13b}.
Note that the 1-bit~DAC precoding problem can be seen as a special (or extreme) case of constant-envelope precoding, where the phase of the transmitted signal is limited to only four different values.

\subsection{Contributions}
We consider quantized precoding for the massive MU-MIMO downlink over frequency-flat channels. Similarly to \cite{usman16a, guerreiro16a, saxena16a}, we consider DACs operating at symbol rate sampling frequency.
However, in contrast to~\cite{usman16a, guerreiro16a, saxena16a}, we do not restrict ourselves to 1-bit~DACs and linear precoding. Specifically, we consider both \emph{linear-quantized precoders}, where a linear precoder is followed by a finite-resolution DAC, and \emph{nonlinear precoders}, where the data vector together with the CSI is used to directly generate the DAC outputs.
Our contributions can be summarized as follows. 
\begin{itemize}
\item We formulate the MMSE-optimal linear-quantized precoding problem and present low complexity, suboptimal linear-quantized precoders that yield approximate solutions to this problem. 
We use Bussgang's theorem to develop simple closed-form approximations for the rate achievable with linear-quantized precoding and low-resolution DACs. 
Through numerical simulations, we validate the accuracy of these approximations, and we show that only a small number of quantization bits are sufficient to close the performance gap to the infinite-resolution case.
For the special case of~$1$-bit DACs, we obtain a firm lower bound on the achievable rate with linear precoding.
\item For the $1$-bit case, we develop a variety of low-complexity nonlinear precoders that achieve near-optimal performance. We show that the MMSE-optimal downlink precoding problem can be relaxed to a convex problem that can be solved in a computationally-efficient manner.
We propose computationally efficient algorithms based on semidefinite relaxation, squared-$\ell_\infty$ norm relaxation, and sphere decoding, and discuss advantages and limitations of each of these methods.
Through numerical simulations, we demonstrate the superiority of nonlinear precoding over linear-quantized precoding.
\item We investigate the sensitivity of the proposed precoders to channel-estimation errors and demonstrate that the proposed precoders are robust to imperfect CSI at the~BS.
\end{itemize}

Our results reveal that massive MU-MIMO enables the use of low-resolution DACs at the BS without a significant performance loss in terms of error-rate performance and information-theoretic rates.

\begin{figure*}[t!]
\centering
 \includegraphics[width=.7\textwidth]{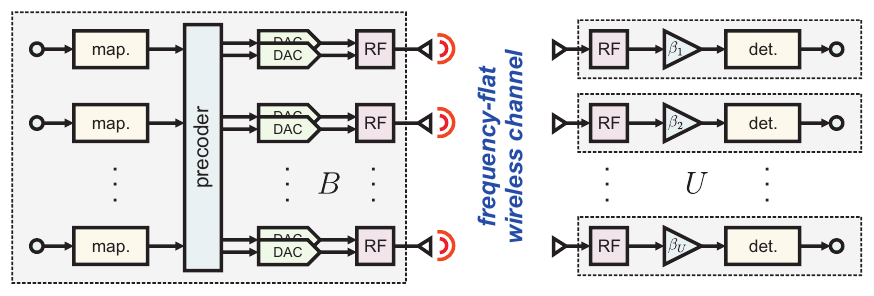}
 \caption{Overview of the proposed quantized massive MU-MIMO downlink system. Left:  $B$ antenna massive MU-MIMO BS that performs quantized~precoding to enable the use of low-resolution DACs; Right: $U$ single-antenna~UEs. } 
  \label{fig:overview}
\end{figure*}

\subsection{Notation}

Lowercase and uppercase boldface letters designate column vectors and matrices, respectively. 
For a matrix $\bA$, we denote its complex conjugate, transpose, and Hermitian transpose by $\bA^*$,~$\bA^T$, and~$\bA^H$, respectively. The entry on the $k$th row and on the $\ell$th column of the matrix $\matA$ is denoted as $[\matA]_{k,\ell}$.
For a vector $\veca$, the $k$th entry is denoted as~$[\veca]_k$.
We use $\matA \succeq \matzero$ to indicate that the matrix~$\matA$ is positive semidefinite.
The trace and the main diagonal of~$\bA$ are~$\tr(\bA)$ and~$\diag(\bA)$, respectively.
The $M\times M$ identity matrix and the all-zeros matrix are denoted by $\bI_M$ and~$\mathbf{0}_{M\times N}$, respectively.
The real and imaginary parts of a complex vector $\veca$ are $\Re\{\veca\}$ and $\Im\{\veca\}$, respectively. 
We use~$\vecnorm{\veca}_2$ and $\vecnorm{\veca}_\infty$ to denote the $\ell_2$-norm and the~$\ell_\infty$-norm of $\veca$, respectively.
We use~$\text{sgn}(\cdot)$ to denote the signum function, which is applied entry-wise to vectors and defined as~$\text{sgn}(a)=+1$ if~$a\ge0$ and~$\text{sgn}(a)=-1$ if~$a<0$.
We further use $\mathds{1}_{\!\setA}(a)$ to denote the indicator function, which is defined as~$\mathds{1}_{\!\setA}(a) = 1$ for $a \in \setA$ and $\mathds{1}_{\!\setA}(a) = 0$ for~$a \notin \setA$.
The multivariate complex-valued circularly-symmetric Gaussian probability density function (PDF) with  covariance matrix~$\bK$ is denoted by~$\setC\setN(\bZero,\bK)$. We use~$f(\cdot)$ to denote PDFs and~$\Ex{\vecx}{\cdot}$ to denote expectation with respect to the random vector~$\vecx$.
The mutual information between two random vectors $\vecx$ and $\vecy$ is written as~$\setI(\vecx;\vecy)$.

\subsection{Paper Outline}

The rest of the paper is organized as follows. 
In \fref{sec:systemmodel}, we introduce the system model and formulate the MMSE-optimal quantized precoding problem.
In \fref{sec:linearprecoders}, we investigate linear-quantized precoders for massive MU-MIMO systems. \fref{sec:nonlinearprecoders} deals with nonlinear precoding algorithms for the case of 1-bit~DACs.
In \fref{sec:simulation}, we provide numerical simulation results and we analyze the robustness of the developed algorithms to channel-estimation errors. 
We conclude the paper in \fref{sec:conclusions}.

\section{System Model and Quantized Precoding}
\label{sec:systemmodel}


\subsection{System Model} \label{sec:system}

We consider the downlink of a single-cell  massive MU-MIMO system as illustrated in \fref{fig:overview}. 
{The system consists of a BS with $B$ antennas that serves $U$ single-antenna UEs simultaneously and in the same time-frequency resource. 
For simplicity, we assume that all RF hardware (e.g., local oscillators, mixers, power amplifiers, etc.) are ideal and that the ADCs at the UEs have infinite resolution. We also assume that the sampling rate of the DACs at the BS is equal to the sampling rate of the ADCs at the UEs and that the system is perfectly synchronized. Finally, we assume that the \emph{reconstruction stage} (see, e.g.,~\cite{maloberti07a}) of the DACs consists only of a zero-order hold circuit (no filtering stage).\footnote{Symbol-rate sampling combined with low-resolution DACs may yield undesired out-of-band emissions, which may be mitigated by using analog filters. Such filters, however, may in turn cause inter-symbol-interference. In this work, we shall ignore the out-of-band emissions caused by the low-resolution DACs and no filter will be considered.}
Under these assumptions, the input-output relation of the downlink channel can be modeled~as
\begin{align} \label{eq:complex_channel}
\vecy = \matH \vecx + \vecn.
\end{align}
Here, the vector $\vecy = [y_1,\,\dots,\,y_U]^T$ contains the received signals at all users, with $y_u \in \complexset$ denoting the signal received at the $u$th UE. 
The matrix $\matH  \in \opC^{U \times B}$ models the downlink channel, and it is assumed to be perfectly known to the BS.\footnote{In~\fref{sec:imperfectCSI}, we will relax this assumption by investigating the impact of imperfect CSI to the robustness of the proposed quantized precoding algorithms.} We shall also assume that the entries of $\matH$ are independent circularly-symmetric complex Gaussian random variables with unit variance, i.e., $h_{u,b} = [\matH]_{u,b}\sim\jpg(0,1)$, for~$u = 1,\,\dots,\,U$, and~$b = 1,\,\dots,\,B$. 
The vector $\bmn\in\complexset^U$ in~\eqref{eq:complex_channel} models additive noise. We assume the noise to be i.i.d.\ circularly-symmetric complex Gaussian with variance $N_0$ per complex entry, i.e.,~$n_u\sim \jpg(0,N_0)$, for~$u = 1,\,\dots,\,U$. We shall also assume that the noise level is known perfectly at the~BS.\footnote{Knowledge of $N_0$ at the BS can be obtained by explicit  feedback from the UEs to the BS.}

The precoded vector is denoted by $\bmx\in\setX^B$, where the set~$\setX$ is the transmit alphabet; this set coincides with the set~$\opC$ of complex numbers in the case of infinite-resolution DACs. In real-world BS architectures with finite-resolution DACs, the set~$\setX$ is, however, a finite-cardinality alphabet. 
Specifically, we denote the set of possible real-valued DAC outputs (quantization labels) as $\setL = \{ \ell_0, \dots, \ell_{L-1} \}$. We refer to~$L = \abs{\setL}$ and~$Q = \log_2 L$ as the number of quantization levels and  the number of quantization bits per  real dimension, respectively. For each BS antenna, we assume the same quantization alphabet for the real part and the imaginary part. Hence, the set of  complex-valued DAC outputs at each antenna is $\setX = \setL \times \setL$. Under these assumptions, the $b$th entry of the precoded vector $\vecx$ is $x_b=\ell_R + j\ell_{I} $ where $\ell_R,\ell_I\in\setL$.

\subsection{Precoding} \label{sec:QPP}
%
Let $s_u \in \setO$ for $u=1,\ldots,U$ be the constellation point at the BS intended for the UE $u$; here,~$\setO$ is the set of constellation points (e.g., QPSK).
The BS uses the available CSI, namely the knowledge of the realization of the channel matrix~$\matH$, to precode the symbol vector $\vecs = [s_1,\,\dots,\,s_U]^T$ into a $B$-dimensional precoded vector $\vecx = \precode(\vecs, \matH)$. Here, the function $\precode(\cdot, \cdot): \setO^U \times \opC^{U \times B} \rightarrow \setX^B$ represents the precoder. The precoded vector $\vecx$ must satisfy the average power constraint 
\begin{IEEEeqnarray}{rCl} \label{eq:powerconstraint}
\Ex{\vecs}{\vecnorm{\vecx}^2_2} \le P.
\end{IEEEeqnarray}
We define $\snr=P/\No$ as the SNR.

Coherent transmission of data using multiple BS antennas leads to an \emph{array gain}, which depends on the realization of the fading channel. 
We shall assume that the $u$th UE is able to rescale the received signal $y_u$ by a factor $\beta_u \in \opR$ to compute an estimate $\hat s_u \in \opC$ of the transmitted symbol $s_u \in \setO$ as~follows:
\begin{IEEEeqnarray}{rCl} \label{eq:scaling}
\hat{s}_u = \beta_u y_u	.
\end{IEEEeqnarray}
The problem of downlink precoding has been studied extensively in the literature. Broadly speaking, the goal is to increase the array gain to the intended UE while simultaneously reducing MU interference~(MUI)~\cite{bjornson14b}. There exist multiple formulations of this optimization problem based on different performance metrics (e.g.,~sum-rate throughput, worst-case throughput, error probability, etc.). We refer the interested reader to the tutorial~\cite{bjornson13a} for a comprehensive overview.

Our specific goal is to design a precoder that minimizes the MSE between the received signal and the transmitted symbol vector $\vecs$ under the power constraint~\eqref{eq:powerconstraint}.
This problem has been studied extensively for the case of infinite-resolution DACs (see, e.g., \cite{christensen08a, joham05a, shi07a}). 
If the BS is equipped with finite-resolution DACs, then the UEs will experience additional distortion compared to the infinite-resolution case, due to finite cardinality of the set~$\setX^B$ of possible precoder outputs.

\begin{figure}[!t]
\centering
\subfloat[Linear-quantized precoders: the precoding matrix $\matP$ is designed based on $\matH$. The transmit vector is the quantized version of $\matP\vecs$, i.e., $\vecx = \quantize(\matP\vecs)$. Here,~$\quantize(\cdot)$ denotes the~quantizer.]{\includegraphics[width = .4\textwidth]{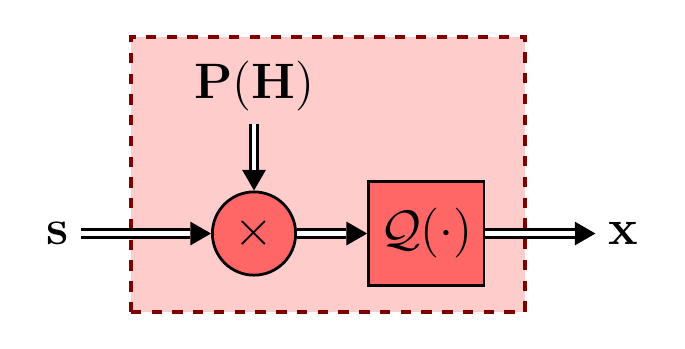}\label{fig:precoder_linear}}
\\
\subfloat[Nonlinear precoders: the quantized transmit vector $\vecx\in\setX^B$ is a nonlinear function of~$\vecs$ and~$\matH$, i.e.,~$\vecx = \precode(\vecs, \matH)$.]{
\includegraphics[width = .4\textwidth]{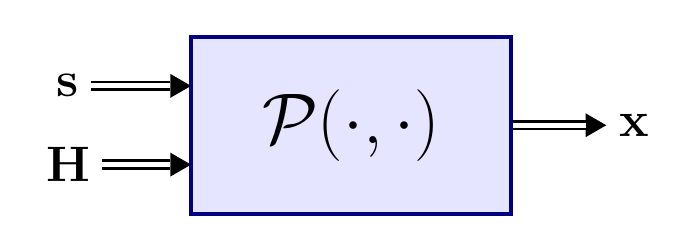}\label{fig:precoder_nonlinear}}
\caption{Illustration of linear-quantized (a) and nonlinear (b) precoders.} 
\label{fig:precoders}
\end{figure}

Finding the MMSE-optimal precoder for BS architectures with finite-resolution DACs is a formidable task due to the finite cardinality of $\setX^B$.
In what follows, we present novel algorithms that efficiently compute approximate solutions to the quantized precoding problem. More specifically, we investigate two approaches: \emph{linear-quantized precoding} in \fref{sec:linearprecoders} and \emph{nonlinear-quantized precoding} for the special case of 1-bit DACs in \fref{sec:nonlinearprecoders}. 
As illustrated in \fref{fig:precoders}, linear-quantized precoders perform linear processing (matrix-vector multiplication) followed by quantization; in contrast, nonlinear precoders use the transmit vector $\bms$ together with the available CSI in order to directly compute the precoded vector~$\bmx$. 
As it will be shown in \fref{sec:simulation}, nonlinear precoders outperform (often significantly) linear-quantized precoders in terms of error-rate performance at the cost of higher computational~complexity.

\section{Linear--Quantized Precoders}
\label{sec:linearprecoders}

In the infinite-resolution case, linear precoders multiply the $U$-dimensional symbol vector $\vecs$ with a precoding matrix $\matP \in \opC^{B \times U}$ so that $\vecx=\matP\vecs$. This approach is particularly attractive for massive MU-MIMO systems due to (i) the relatively low computational complexity and (ii) the fact that even the simplest linear precoder, namely the MRT precoder, achieves virtually optimal performance in the large-antenna limit (see, e.g., \cite{rusek14a}). 
Linear-quantized precoders inherit the first of these two advantages. Indeed, quantizing the precoded vector implies no additional computational complexity. 
For linear-quantized precoders, the precoded vector $\vecx \in \setX^B$ is~given~by
\begin{align} \label{eq:linear_quantize}
\vecx  = \quantize(\matP\vecs).
\end{align}
Here, $\quantize(\cdot): \opC^B \rightarrow \setX^B$ denotes the \emph{quantizer-mapping} function, which is a nonlinear function that describes the joint operation of the $2B$ DACs at the BS.

The remainder of this section is organized as follows.
We start by formulating the MMSE quantized precoding problem for linear-quantized precoders.
We then describe the operation of the DACs and define the quantizer-mapping function. 
We then use Bussgang's theorem \cite{bussgang52a} to derive a lower bound on the sum-rate capacity for the case of $1$-bit DACs at the BS.
Finally,  we derive a simple closed-form approximation of the rate achievable with Gaussian inputs for the more general case of $Q$-bit DACs.

\subsection{The Linear-Quantized Precoding Problem} \label{sec:LQPP}
%
By restricting ourselves to linear-quantized precoding~(LQP), we can formulate the quantized precoding problem as follows:
\begin{IEEEeqnarray}{rCl} \label{eq:problem_linear}
&&\!\!\!\text{(LQP)} \ \left\{\begin{array}{cll}
\!\!\underset{\matP \in \opC^{B \times U}\!, \beta \in \reals}{\text{minimize}} & \!\!\! \Ex{\vecs}{\vecnorm{\vecs -\! \beta \matH\quantize(\matP\vecs)}^2_2} \! + \! \beta^2 U N_0\\
\!\!\text{subject to} & \!\!\! \Ex{\vecs}{\vecnorm{\vecx}^2_2} \le P \text{ and } \beta > 0.
\end{array}\right. 
\end{IEEEeqnarray}
The resulting precoding matrix $\matP^\text{LQP}$ and the associated \emph{precoding factor} $\beta^\text{LQP}$ will be referred to as the optimal solution to the problem ($\LQP$). 
Here, we have introduced the scalar $\beta \in \opR$ to account for the array gain at the UEs (as commonly done in the MMSE precoding literature; see, e.g., \cite{joham05a, mezghani09c}).
By solving \eqref{eq:problem_linear}, we find the precoded vector $\vecx^\text{LQP}$ that minimizes the per-channel MSE between the transmitted symbols $\vecs$ and the vector $\beta \vecy$.
Indeed, note that
\begin{IEEEeqnarray}{rCl} \label{eq:mse}
\Ex{\vecs}{\vecnorm{\vecs - \beta\vecy}_2^2} &=& \Ex{\vecs}{\vecnorm{\vecs - \beta\matH\vecx}_2^2} + \beta^2UN_0 \IEEEeqnarraynumspace
\end{IEEEeqnarray}
and recall that $\vecx = \quantize(\matP\vecs)$. 
Next, we provide more insights on the role of the precoding factor $\beta$. We seek a precoded vector~$\vecx$ that makes the received signal proportional to the transmitted symbol vector~$\vecs$, i.e., $\vecs \approx \beta\vecy$. To lessen the adverse impact of the noise vector $\vecn$ in \eqref{eq:complex_channel}, we look for a design that maximizes the received signal power at the UEs. The cost function in~\eqref{eq:problem_linear} accomplishes exactly this goal by favoring solutions with a smaller $\beta$.
Unfortunately, the introduction of the precoding factor $\beta$ (which is not known to the UEs) may complicate decoding at the UEs.\footnote{We shall elaborate on this point in \fref{sec:uedecoding}.}

%
%

Solving~\eqref{eq:problem_linear} in closed form is challenging due to the nonlinear operation of the DACs, which is captured by the quantizer-mapping function $\quantize(\cdot)$. An approximate solution to~\eqref{eq:problem_linear} was given in \cite{mezghani09c}. This solution is obtained by approximating the statistics of the distortion caused by the DACs.
We shall consider here a different approach. Specifically, we design linear precoders that assume infinite-resolution DACs at the BS, and then quantize the resulting precoded vector. Such linear-quantized precoders have the advantage that the precoding matrix $\matP$ does not depend on the resolution of the DACs.
Furthermore, as we shall see in~\fref{sec:errorrate}, the difference in error-rate performance between the precoders found using our approach and the precoder presented in~\cite{mezghani09c} is negligible. 
We next review a selection of linear precoding algorithms for the case of infinite-resolution DACs.

\subsubsection{WF precoding}

For the case when the BS is equipped with infinite-resolution DACs, the solution to~\eqref{eq:problem_linear} is the Wiener filter (WF) precoder~\cite{joham05a}:
\begin{IEEEeqnarray}{rCl} \label{eq:P_WF}
\matP^\text{WF} = \frac{1}{\beta^\text{WF}}\matH^H \lefto(\matH\matH^H + \frac{UN_0}{P}\matI_U\right)^{-1}
\end{IEEEeqnarray}
where
\begin{IEEEeqnarray}{rCl}
	\beta^\text{WF} &=& \frac{1}{\sqrt{P}} \tr\lefto( \lefto(\matH\matH^H + \frac{UN_0}{P}\matI_U\right)^{-1} \right. \nonumber \\ && \left. \matH\matH^H\lefto(\matH\matH^H + \frac{UN_0}{P}\matI_U\right)^{-1} \right)^{-1/2}.  \label{eq:beta_WF}
\end{IEEEeqnarray}
%
%
We write the resulting precoded vector as $\vecx^\text{WF} = \quantize\lefto(\matP^\text{WF}\vecs\right)$.

\subsubsection{ZF precoding}

With ZF precoding, the BS nulls the MUI by choosing as precoding matrix the pseudoinverse of the channel matrix. The ZF precoding matrix is obtained from \eqref{eq:P_WF} by setting the noise variance $N_0$ to zero, which yields $\matP^\text{ZF} = \frac{1}{\beta^\text{ZF}}\matH^\dagger$, where $\matH^\dagger = \matH^H(\matH\matH^H)^{-1}$ is the pseudoinverse of the channel matrix~$\matH$, and $\beta^\text{ZF} = \frac{1}{\sqrt{P}} \sqrt{\tr\lefto((\matH\matH^H)^{-1} \right)}$. The resulting precoded vector is~$\vecx^\text{ZF} = \quantize\lefto(\matP^\text{ZF}\vecs\right)$.

\subsubsection{MRT precoding}

The MRT precoder maximizes the power directed towards each UE, ignoring MUI.
The precoding matrix can be obtained from \eqref{eq:P_WF} by letting the noise variance $N_0$~tend to infinity, which yields $\matP^\text{MRT} = \frac{1}{\beta^\text{MRT}B} \matH^H$ and~$\beta^\text{MRT} = \frac{1}{B\sqrt{P}}\sqrt{\tr\lefto(\matH\matH^H \right)}$. The resulting precoded vector is~$\vecx^\text{MRT} = \quantize\lefto(\matP^\text{MRT}\vecs\right)$.

\subsection{Uniform Quantization of a Complex-Valued Vector}

For simplicity, we shall model the DACs as symmetric uniform quantizers with step size $\Delta$. When a signal is quantized, the average power in the signal is in general not preserved. Therefore, we further assume that the output of the quantizer is scaled by a constant $\alpha \in \opR$, to ensure that the transmit power constraint \eqref{eq:powerconstraint} is satisfied.
We start by defining a set of quantization labels $\setL = \{\ell_0, \dots, \ell_{L-1} \}$ with entries
\begin{IEEEeqnarray}{rCl} \label{eq:labels}
\ell_i &=& \alpha\Delta \lefto( i - \frac{L-1}{2}\right)\!, \quad i=0,\ldots,L-1.
\end{IEEEeqnarray}
Furthermore, let $\setT = \{\tau_0, \dots, \tau_{L} \}$, where $-\infty = \tau_0 < \tau_1 < \dots < \tau_{L-1} <\tau_{L} = \infty$ specify the set of~$L+1$ quantization thresholds. For uniform quantizers, the quantization thresholds are given by
\begin{IEEEeqnarray}{rCl} \label{eq:thresholds}
\tau_i &=& \Delta \lefto( i - \frac{L}{2}\right)\!, \quad i=1,\ldots,L-1.
\end{IEEEeqnarray}
The quantizer-mapping function $\quantize(\cdot)$ can be uniquely described by the set of quantization labels~$\setL$ and the set of quantization thresholds~$\setT$. The DACs map $\vecz \in \opC$ with entries $\{ z_{b}\}$ into the quantized output $\vecx$ with entries $\{ x_{b} \}$ in the following way: if $\Re\{ z_{b} \} \in [\tau_k, \tau_{k+1})$ and $\Im\{ z_{b} \} \in [\tau_l, \tau_{l+1})$, then $x_{b} = \ell_k + j \ell_l$. 

The step size $\Delta$ of the quantizers should be chosen to minimize the distortion between the quantized and nonquantized vector. The optimal step size $\Delta$ depends on the distribution of the input~\cite{hui01a}, which in our case depends on both the precoder and the signaling scheme. For simplicity, we set the step size so as to minimize the distortion under the assumption that the per-antenna input to the quantizers is $\jpg(0,P/B)$-distributed. This step size can be found numerically (see e.g., \cite{al-dhahir96a} for~details).

In the extreme case of 1-bit DACs, the quantizer-mapping function reduces to
\begin{IEEEeqnarray}{rCl} \label{eq:quantizer_1bit}
\quantize(\vecz) = \sqrt{\frac{P}{2B}} \big(\! \sign(\Re\{ \vecz \}) + j\sign(\Im\{ \vecz \})\big).
\end{IEEEeqnarray}
Here, we have chosen the set of possible complex-valued quantization outcomes per antenna to be~$\setX=\{\sqrt{P/(2B)}\,\lefto(\pm 1 \pm j\right)\}$, which ensures that the power constraint in~\eqref{eq:powerconstraint} is satisfied with~equality.

\subsection{Signal Decomposition using Bussgang's Theorem}

Quantizing the precoded signal causes a distortion $\quantize(\matP\vecs) - \matP\vecs$ that is correlated with the input $\matP\vecs$ to the DACs.  
For Gaussian inputs, Bussgang's theorem~\cite{bussgang52a} allows us to decompose the quantized signal into a linear function of the input to the quantizers and a distortion term that is \emph{uncorrelated} with the input to the quantizers~\cite{rowe82a, zhang16a}. This allows us to characterize the rates achievable with Gaussian inputs. We start by stating Bussgang's theorem~\cite{bussgang52a,rowe82a}.

\begin{thm}
\label{thm:bussgang} Consider two zero-mean jointly complex Gaussian random variables~$x$ and~$y$. Assume that~$x$ is passed through a nonlinear function~$g(\cdot): \opC \rightarrow \opC$ that acts independently on the real and the imaginary components of $x$. The covariance between $g(x)$ and $y$ is given by
\begin{IEEEeqnarray}{rCl} \label{eq:bussgang_complex}
\Ex{}{g(x)y^*} &=& \frac{\Ex{}{g(x)x^*}}{\Ex{}{xx^*}} \Ex{}{xy^*}.
\end{IEEEeqnarray}
%
\end{thm}

Bussgang's theorem has recently been used to analyze the massive MU-MIMO uplink with $1$-bit ADCs (see, e.g., \cite{li17b, mollen16c}). It has also been used in \cite{guerreiro16a} to approximate the distortion levels caused by MRT precoding and $1$-bit quantization in the massive MIMO downlink.
We shall use \fref{thm:bussgang} to characterize the performance of linear-quantized precoders for the case of $Q$-bit uniform DACs. As a first step, we establish the following result, whose proof is given in \fref{app:appB}.

%
\begin{thm} \label{thm:decomp_uniform}
%
Let $\vecx = \quantize(\matP\vecs)$ denote the output from a set of uniform quantizers described by the quantizer-mapping function $\quantize: \opC^B \rightarrow \setX^B$. Assume that~$\matP \in \opC^{B \times U}$ and that $\vecs \sim \jpg(\veczero, \matI_U)$. The quantized vector $\vecx$ can be decomposed as 
\begin{IEEEeqnarray}{rCl} \label{eq:x_decomp}
\vecx = \matG\matP\vecs + \vecd
\end{IEEEeqnarray}
where the distortion $\vecd$ and the signal $\vecs$ are uncorrelated. Furthermore, $\matG \in \opR^{B \times B}$ is the following diagonal matrix:
\begin{IEEEeqnarray}{rCl}
\matG&=& \frac{\alpha\Delta}{\sqrt{\pi}}  \diag\lefto(\matP\matP^H\right)^{-1/2} \nonumber\\  
&&\sum_{i=1}^{L-1} \exp\lefto(-\Delta^2\lefto( i - \frac{L}{2} \right)^2\diag\lefto(\matP\matP^H\right)^{-1}\right).  \label{eq:gainmatrix_uniform}
\end{IEEEeqnarray}
Here, $L$ and $\Delta$ denote the number of levels and  the step size of the DACs, respectively.
\end{thm}

The following corollary provides a well-known result for the case of 1-bit quantization (see, e.g.,~\cite{li17b,mollen16c}). Its proof follows by setting~$L=2$ and~$\alpha\Delta = \sqrt{2P/B}$ in~\eqref{eq:gainmatrix_uniform} to satisfy the power constraint~\eqref{eq:powerconstraint} with equality.
\begin{cor} \label{cor:decomp_1bit} For the case of 1-bit DACs, the matrix~$\matG$ in~\eqref{eq:gainmatrix_uniform} reduces to
\begin{IEEEeqnarray}{rCl} \label{eq:gainmatrix_1bit}
\matG &=& \sqrt{\frac{2P}{\pi B}} \,\diag\lefto(\matP\matP^H\right)^{-1/2}.
\end{IEEEeqnarray}
\end{cor}

Let now $\vech_u^T$ denote the $u$th row of the channel matrix $\matH$, let $\vecp_u$ be the $u$th column of the precoding matrix~$\matP$, and $n_u$ be the $u$th entry of the noise vector $\vecn$.
Using \eqref{eq:x_decomp}, we can express the received signal $y_u$ at UE $u$ as~follows:
\begin{IEEEeqnarray}{rCl} \label{eq:y_u_decomp}
y_u &=& \vech_u^T\matG\matP\vecs + n_u \\[7pt]
&=& \vech_u^T\matG\vecp_us_u + \sum_{v \neq u} \vech_u^T\matG\vecp_vs_v  + \vech_u^T\vecd + n_u \\[-1pt]
&=& \vech_u^T\matG\vecp_us_u + e_u + n_u. \IEEEeqnarraynumspace
\end{IEEEeqnarray}
Here, the error term $e_u = \sum_{v \neq u} \vech_u^T\matG\vecp_vs_v  + \vech_u^T\vecd$ captures both the MUI and the distortion caused by the finite-resolution DACs. Note that $e_u$ and $s_u$ are uncorrelated. Indeed, 
\begin{IEEEeqnarray}{rCl}
\Ex{\vecs}{e_u s_u^*}
&=& \textstyle \sum_{v \neq u} \vech_u^T\matG\vecp_v \Ex{\vecs}{s_vs_u^*} + \vech_u^T\Ex{\vecs}{\vecd s_u^*} = 0. \IEEEeqnarraynumspace
\end{IEEEeqnarray}
We shall next use the decomposition in~\eqref{eq:y_u_decomp} to analyze the performance of linear-quantized precoders.

%

\subsection{Achievable Rate Lower Bound for 1-bit DACs}
\label{sec:achrate}

We assume that each UE scales its received signal $y_u$ by the scalar $\beta_u = (\vech_u^T\matG\vecp_u)^{-1}$ (which is assumed to be known at the $u$th UE) to obtain the following estimate:
\begin{IEEEeqnarray}{rCl} \label{eq:rx_linear_decomposed}
\hat{s}_u &=& \beta_uy_u = s_u + \beta_u(e_u + {n}_u).
\end{IEEEeqnarray}
The nonlinearity introduced by the DACs prevents one to characterize the probability distribution of the error term $e_u$ in closed form, which  makes it difficult to compute the achievable rates.
One can, however, lower-bound the achievable rate using the so-called ``auxiliary-channel lower bound'' \cite[p.~3503]{arnold06a}, which gives the rates achievable with a mismatched decoder (see~\cite[ch.\,$1$]{scarlett14a} for a recent review on the subject). As auxiliary channel, we take the one with output  
\begin{IEEEeqnarray}{rCl} \label{eq:aux_channel}
\tilde{s}_u &=& s_u + \beta_u(\tilde{e}_u + {n}_u),
\end{IEEEeqnarray}
where $\tilde{e}_u \sim \jpg\lefto(0,\Ex{\vecs}{|e_u|^2}\right)$ has the same variance as the actual error term $e_u$ but is Gaussian distributed.
Assuming Gaussian inputs, by standard manipulations of the mutual information, we can bound the achievable rate $R_u$ for UE $u = 1,2,\dots,U$ as follows:
\begin{IEEEeqnarray}{rCl} 
R_u &=& \Ex{\matH}{\,\setI\lefto(s_u; \hat{s}_u \given \matH \right)}\\[5pt]
	&=& \Ex{s_u, \hat{s}_u, \matH}{\log_2\lefto(\frac{f_{\hat{s}_u | s_u, \matH}(\hat{s}_u | s_u, \matH)}{f_{\hat{s}_u\given\matH}(\hat{s}_u\given\matH)}\right)} \label{eq:Ru_def}\\
	&\ge& \Ex{s_u, \hat{s}_u, \matH}{\log_2\lefto(\frac{f_{\tilde{s}_u | s_u, \matH}(\hat{s}_u | s_u, \matH)}{f_{\tilde{s}_u\given\matH}(\hat{s}_u\given\matH)}\right)}\\[5pt] 
	&=& \Ex{\matH}{\log_2(1 + \sindr_u)} \label{eq:rate_lower_bound}
\end{IEEEeqnarray}
where
\begin{IEEEeqnarray}{rCl} \label{eq:sindr_linear}
\sindr_u
&=& \frac{\abs{\vech_u^T\matG\vecp_u}^2}{\sum_{v \neq u}\abs{\vech_u^T\matG\vecp_v}^2 + \vech_u^T \matC_{\vecd\vecd}\vech_u^* + N_0}
\end{IEEEeqnarray}
is the signal-to-interference-noise-and-distortion ratio (SINDR) at the $u$th UE.\footnote{One can establish~\eqref{eq:rate_lower_bound} also by noting that Gaussian noise is the worst noise for Gaussian inputs~\cite{lapidoth96a}.} 
Here, $\matC_{\vecd\vecd} = \Ex{\vecs}{\vecd\vecd^H}$ denotes the covariance of the distortion $\vecd$.
It is worth pointing out that the choice of the auxiliary channel~\eqref{eq:aux_channel} corresponds to the use of mismatched nearest-neighbor decoding at the UEs~\cite{lapidoth96b, zhang12a}.

Next, we use~\eqref{eq:x_decomp} to write the covariance matrix $\matC_{\vecd\vecd}$ in~\eqref{eq:sindr_linear}~as
\begin{IEEEeqnarray}{rCl} \label{eq:distcovariance}
\matC_{\vecd\vecd} = \matC_{\vecx\vecx} - \matG\matP\matP^H\matG^H
\end{IEEEeqnarray}
where $\matC_{\vecx\vecx} = \Ex{\vecs}{\vecx\vecx^H}$ is the covariance matrix of the quantized signal $\vecx = \quantize(\matP\vecs)$. In the special case of 1-bit~DACs,~$\matC_{\vecx\vecx}$ can be written in closed-form as~\cite{van-vleck66a,jacovitti94a}
\begin{IEEEeqnarray}{rCl} \label{eq:arcsine}
 &&\matC_{\vecx\vecx} \!=\! \frac{P}{\pi B} \lefto(\sin^{-1}\lefto(\!\diag(\matP\matP^H)^{-\frac{1}{2}} \Re\{ \matP\matP^H \} \diag(\matP\matP^H)^{-\frac{1}{2}}\right)\right. \nonumber\\ 
&& \quad\! +\left. j\sin^{-1}\lefto( \diag(\matP\matP^H)^{-\frac{1}{2}} \Im\{ \matP\matP^H \} \diag(\matP\matP^H)^{-\frac{1}{2}}\right)\right.\!.
\end{IEEEeqnarray}
Thus, using~\eqref{eq:distcovariance} and \eqref{eq:arcsine}, we can express the SINDR in \eqref{eq:sindr_linear} in closed form for the case of 1-bit DACs. Substituting~\eqref{eq:sindr_linear} in~\eqref{eq:rate_lower_bound}, one obtains a lower bound on the per-user achievable rate with Gaussian signaling for the 1-bit DAC case. Unfortunately, no closed-form expression for $\matC_{\vecx\vecx}$ is available for the multi-bit DAC case. We address this problem in the next~section.

\subsection{Achievable Rate Approximation for Multi-Bit~DACs}

In this section, we provide an approximation of \eqref{eq:sindr_linear} for the multi-bit DAC case, which is derived under the assumption that both $B$ and $U$ are large and that the error term $e_u$ in \eqref{eq:y_u_decomp} is a Gaussian random variable. The approximation relies on standard random matrix theory arguments. Specifically, let
\begin{IEEEeqnarray}{rCl} \label{eq:gain_avg}
G &=& \alpha\Delta \sqrt{\frac{B}{\pi P}} \sum_{i=1}^{L-1} \exp\lefto(-\frac{B\Delta^2}{P}\lefto( i - \frac{L}{2} \right)^2\right)
\end{IEEEeqnarray}
where the normalization by $\alpha$, given by
\begin{IEEEeqnarray}{rCl} \label{eq:alpha_gauss}
\alpha &=& \lefto(2B\Delta^2 \lefto( \lefto(\frac{L-1}{2}\right)^2 \right.\right. \nonumber\\
&& \lefto.\lefto. - 2\sum_{i=1}^{L-1} \lefto( i - \frac{L}{2}\right) \Phi\lefto( \sqrt{2B\Delta^2}\lefto( i - \frac{L}{2}\right)\right)\right)\right)^{-1/2} \IEEEeqnarraynumspace
\end{IEEEeqnarray}
ensures that the power constraint \eqref{eq:powerconstraint} is satisfied. In~\eqref{eq:gain_avg}, the function $\Phi(x) = \int_{-\infty}^x \frac{1}{\sqrt{2\pi}}\,e^{-{t^2}/{2}}\,dt$ is the cumulative distribution function of a Gaussian random variable.
Let also $\bar{\snr}$ be defined as follows:
\begin{IEEEeqnarray}{rCl} \label{eq:SNR_eff}
\bar\snr &=& \frac{G^2\snr}{(1-G^2)\snr + 1}.
\end{IEEEeqnarray}
Following the same approach as in~\cite{coulliet11a, wagner12a, hoydis13a}, one can show that, for the three linear-quantized precoders (WF, ZF, and MRT) introduced in~\fref{sec:LQPP}, the SINDR~$\gamma_u$ in~\eqref{eq:sindr_linear} can be approximated for large $B$ and $U$ by
\begin{IEEEeqnarray}{rCl} 
&&\bar\sindr^\text{WF} = \frac{\bar\snr}{2}\lefto(\frac{B}{U}-1\right) - \frac{1}{2} \nonumber\\  
&&\quad\quad\,\,\,\ +\,\frac{1}{2}\sqrt{ \bar\snr^2\lefto(\frac{B}{U}-1\right)^2 + 2\bar\snr\lefto(\frac{B}{U}+1\right) + 1} \label{eq:SINDR_WF}\\
&&\bar\sindr^\text{ZF} = \bar\snr\lefto(\frac{B}{U}-1\right) \label{eq:SINDR_ZF}\\
&&\bar\sindr^\text{MRT} = \frac{\bar\snr B}{\bar\snr(U-1) + U}. \label{eq:SINDR_MRT}
\end{IEEEeqnarray}
Substituting~\eqref{eq:SINDR_WF}--\eqref{eq:SINDR_MRT} into~\eqref{eq:rate_lower_bound}, one gets an approximation of the achievable rate with Gaussian signaling and nearest-neighbor decoding that is valid for large $B$ and $U$.
In~\fref{sec:simulation}, we verify through numerical simulations that this approximations is accurate already for realistic values of $B$ and $U$.

\section{Nonlinear Precoders for 1-Bit DACs}
\label{sec:nonlinearprecoders}

We now investigate nonlinear precoders that seek approximate solutions to the MMSE-optimal problem detailed in \fref{sec:QPP}.
We shall focus on the extreme case of 1-bit DACs, for which the problem simplifies and efficient numerical algorithms can be developed.

We start by noting that, in the 1-bit case, all DAC outcomes have equal amplitude, and that $\vecnorm{\vecx}_2^2 = P$ if one sets $\alpha\Delta = \sqrt{2P/B}$ in \eqref{eq:labels}. This observation allows us to formulate the 1-bit quantized precoding~($\OBQP$) problem as follows:
\begin{align} \label{eq:problem_complex_1bit}
\text{(\OBQP)} \quad \left\{\begin{array}{lll}
\underset{\bmx \in \setX^B,\, \beta \in \reals}{\text{minimize}} & \vecnorm{\vecs - \beta \matH\vecx}^2_2 + \beta^2 U N_0\\
\text{subject to} & \beta > 0.
\end{array}\right.
\end{align}
Here, $\setX = \big\{\sqrt{P/(2B)}\,\lefto(\pm 1 \pm j\right)\big\}$. The resulting precoded vector $\vecx^\text{QP}$ and the associated precoding factor~$\beta^\text{QP}$ are referred to as the optimal solution to the problem \eqref{eq:problem_complex_1bit}. 

Compared to the problem (LQP) in~\eqref{eq:problem_linear}, where we minimize the MSE averaged over both the symbol vector~$\vecs$ and the noise vector $\vecn$ (for a given $\matH$), in (QP) we minimize the MSE averaged over the noise vector~$\vecn$ (for a given $\vecs$ and $\matH$). 
Since the optimization problem is solved for a given~$\vecs$, the precoding factor~$\beta$ depends on~$\vecs$; this is in contrast to the linear-quantized case, where $\beta$ depends only on $\matH$.\footnote{We shall discuss how the dependence of $\beta$ on $\vecs$ effects decoding the receiver side in~\fref{sec:uedecoding}.}

We note that (QP) in \fref{eq:problem_complex_1bit} resembles an $\ell_2$-norm regularized closest-vector problem (CVP), with the unique feature that the discrete set of vectors is parametrized by the continuous precoding factor~$\beta$. This prevents the straightforward use of conventional algorithms to approximate CVPs \cite{agrell02a,fincke85a}. 
Since the objective function in \eqref{eq:problem_complex_1bit} is a quadratic function in~$\beta$, we can compute the optimal value of $\beta$ as
\begin{IEEEeqnarray}{rCl} \label{eq:beta_optimal}
\hat\beta(\vecx)
&=& \frac{\Re\{\vecs^H\matH\vecx\}}{\vecnorm{\matH\vecx}_2^2 +U N_0} = 
\frac{\Re\{\vecs^H\matH\vecx\}}{\vecx^H\lefto( \matH^H\matH + \frac{U N_0}{P} \matI_B \right)\vecx} \IEEEeqnarraynumspace
\end{IEEEeqnarray}
which depends on $\vecx$. Inserting~\eqref{eq:beta_optimal} into the objective function in~\eqref{eq:problem_complex_1bit}, we obtain the following equivalent formulation of the QP problem:
\begin{align} \label{eq:problem_complex_beta_optimal}
\begin{array}{lll}
\underset{\bmx \in \setX^B}{\text{minimize}} &  \vecnorm{\vecs - \hat\beta(\vecx) \matH\vecx}^2_2 + \hat\beta(\vecx)^2 U N_0.
\end{array}
\end{align}
To obtain $\beta^\text{\OBQP}$, we can then simply evaluate \fref{eq:beta_optimal} for the optimal vector $\bmx^\text{\OBQP}$. 
We emphasize that a straightforward exhaustive search to solve ($\OBQP$) requires the evaluation of $\abs{\setX}^B = 4^B$ candidate vectors, a quantity that grows exponentially with the number of BS antennas $B$. For a system with $B=128$ antennas at the BS, this approach would require us to evaluate the objective function more than~$10^{77}$ times (more than 10 quattuorvigintillions times). 
In fact, for a fixed value of~$\beta$, the problem~($\OBQP$) is a closest vector problem that is NP-hard \cite{verdu89a}. This implies that there are no known algorithms to solve such problems efficiently for large values of~$B$.\footnote{As we will show in \fref{sec:sphere}, we can---in some cases---design branch-and-bound methods (such as sphere-decoding methods) that allow us to solve the quantized precoding problem efficiently for moderately-sized problems. For massive MU-MIMO systems with hundreds of antennas, however, such methods still exhibit prohibitive computational complexity. }
Hence, alternative algorithms that solve a lower complexity version of the QP problem are required for massive~MU-MIMO systems.

In order to develop such computationally efficient algorithms, we start by defining the auxiliary vector~$\bmb=\beta\bmx$ and rewrite~\eqref{eq:problem_complex_1bit} in the following equivalent form:
\begin{align} \label{eq:problem_tweak_complex}
\underset{\vecb \in \setB^{B}}{\text{minimize}} \quad \vecnorm{\vecs - \matH \vecb}_2^2 + \frac{U N_0}{P} \vecnorm{\vecb}^2_2.
\end{align}
Here, $\setB = \big\{ \sqrt{P/(2B)}\,\lefto(\pm \beta \pm j\beta\right), \text{ for all } \beta > 0 \big\}$. To obtain \eqref{eq:problem_tweak_complex}, we have used that $\beta^2 = \vecnorm{\vecb}_2^2/P$. 
Let $\vecb^\text{QP}$ be the solution to~\eqref{eq:problem_tweak_complex}. The resulting precoding vector is obtained by scaling each entry of $\vecb^\text{QP}$ so that it belongs to the set $\setX$. Clearly, $1/\beta^\text{QP}$ is the scaling parameter.

It turns out convenient to transform the complex-valued problem~\eqref{eq:problem_tweak_complex} into an equivalent real-valued problem using the following definitions:
\begin{IEEEeqnarray}{rCl}
\vecb_\opR &=&\!
\begin{bmatrix}
\Re\{\vecb\} \\ \Im\{\vecb\}
\end{bmatrix}\!,\,
\vecs_\opR =\!
\begin{bmatrix}
\Re\{\vecs\} \\ \Im\{\vecs\}
\end{bmatrix}\!,\,
\text{and }\matH_\opR \!=\! 
\begin{bmatrix}
\Re\{\matH\} & \!\!-\Im\{\matH\} \\ 
\Im\{\matH\} & \!\!\Re\{\matH\}
\end{bmatrix}\!. \nonumber
\end{IEEEeqnarray}
These definitions enable us to rewrite \eqref{eq:problem_tweak_complex} as 
\begin{align} \label{eq:problem_tweak_real}
\underset{\vecb_\opR \in \setB^{2B}_\opR}{\text{minimize}} \quad \vecnorm{\vecs_\opR - \matH_\opR \vecb_\opR}_2^2 + \frac{U N_0}{P} \vecnorm{\vecb_\opR}^2_2
\end{align}
where $\setB_\opR = \big\{ \pm \sqrt{P/(2B)}\,\beta, \text{ for all } \beta > 0 \big\}$ is the set of scaled antipodal outputs of each $1$-bit DAC.
We shall next develop a variety of nonlinear precoding methods that find approximate solutions to the problem~\eqref{eq:problem_tweak_real}.

\subsection{Semidefinite Relaxation}
Semidefinite relaxation (SDR) is a well-established technique to develop approximate algorithms for a variety of discrete programming problems~\cite{luo10a}. For example, SDR is commonly used to find near-ML solutions for the MU-MIMO detection problem (see, e.g., \cite{luo10a, tan01a}).
For the case when the BS is equipped with infinite-resolution DACs, SDR has been used for downlink precoding in \cite{bengtsson99a, sidiropoulos06b}. We next show how SDR can be used to find approximate solutions to~\eqref{eq:problem_complex_1bit}.

In our context, SDR involves relaxing~\eqref{eq:problem_tweak_real} to a semidefinite program (SDP) as follows. We start by writing the real-valued problem~\eqref{eq:problem_tweak_real} in the following equivalent form \cite{luo10a}:
\begin{align} \label{eq:problem_complex_aux}
 \begin{array}{cl}
\underset{\vecb_\opR \in \opR^{2B}, \, \psi \in \{ \pm 1\}}{\text{minimize}} & \vecnorm{\psi\vecs_\opR - \matH_\opR \vecb_\opR}_2^2 + \dfrac{U N_0}{P} \vecnorm{\vecb_\opR}_2^2 \\
\text{subject to} & [\bmb_\opR]_1^2 = [\bmb_\opR]_b ^2, \, b=2,\ldots,2B.
\end{array}
\end{align}
If $\psi = 1$ then $\vecb_\opR$ is the solution to \eqref{eq:problem_complex_aux}; if $\psi = -1$, the solution is $-\vecb_\opR$. 
Next, let the $(2B+1)\times (2B+1)$ matrix $\matT_\opR$ be defined as follows:
\begin{IEEEeqnarray}{rCl}
\matT_\opR &=& 
\begin{bmatrix}
\matH_\opR^T\matH_\opR + \frac{UN_0}{P}\matI_{2B}& -\matH_\opR^T\vecs_\opR \\ 
-\vecs_\opR^T\matH_\opR & \vecnorm{\vecs_\opR}^2_2
\end{bmatrix}\!.
\end{IEEEeqnarray}
Also, let $\matB_\opR = [\vecb_\opR^T \ \psi]^T [\vecb_\opR^T \ \psi]$. Following steps similar to those in \cite{luo10a}, we rewrite the objective function in \eqref{eq:problem_complex_aux} as
\begin{IEEEeqnarray}{rCl} \label{eq:xyzsaft}
\vecnorm{\psi \vecs_\opR - \matH_\opR \vecb_\opR}_2^2 + \frac{U N_0}{P} \vecnorm{\vecb_\opR}_2^2 
= \tr\lefto( \matT_\opR \matB_\opR \right).
\end{IEEEeqnarray}
%
%
The problem~\fref{eq:problem_complex_aux} can now be reformulated~as 
\begin{align} \label{eq:problem_real_1bit_matrix}
\begin{array}{ll}
\underset{\matB_\opR \in \opS^{2B+1}}{\text{minimize}} & \tr\lefto( \matT_\opR \matB_\opR \right) \\
\text{subject to} & [\matB_\opR]_{1,1} = [\matB_\opR]_{b,b} \text{ for } b=2,\ldots,2B, \\
&[\matB_\opR]_{2B+1, \, 2B+1} = 1, \ \matB_\opR \succeq \matzero, \\ 
&\text{ and }\rank(\matB_\opR) = 1.
\end{array}
\end{align}
Here, $\opS^{2B + 1}$ denotes the set of real and symmetric $(2B + 1) \times (2B + 1)$ matrices. 
To see why \eqref{eq:problem_complex_aux} and \eqref{eq:problem_real_1bit_matrix} are equivalent, remember that~$\matB_\opR = [\vecb_\opR^T \ \psi]^T [\vecb_\opR^T \ \psi]$, which implies that $\matB_\opR$ has rank 1, and that~$[\matB_\opR]_{b,b} = [\vecb_\opR]_b^2$ for~$b = 1,\dots, 2B$, and~$[\matB_\opR]_{2B+1,2B+1} = \psi^2 = 1$. 

Unfortunately, the rank-1 constraint in \fref{eq:problem_real_1bit_matrix} is nonconvex, which makes this problem just as hard to solve as the original QP problem in \eqref{eq:problem_complex_1bit}.
Nevertheless, we can use SDR to relax the problem in  \fref{eq:problem_real_1bit_matrix} by omitting the rank-1 constraint, which results in the following SDP:
\begin{IEEEeqnarray}{rCl} \label{eq:problem_sdr_1bit}
&&\!(\SDR)  \left\{
\begin{array}{ll}
\!\!\!\underset{\matB_\opR \in \opS^{2B+1}}{\text{minimize}} & \!\!\!\!\tr\lefto( \matT_\opR \matB_\opR \right) \\
\!\!\!\text{subject to} & \!\!\!\![\matB_\opR]_{1,1} \!=\! [\matB_\opR]_{b,b}, b=\!2,\dots,2B, \\
&\!\!\!\![\matB_\opR]_{2B+1, 2B+1} \!=\! 1,\,\text{and}\,\matB_\opR\!\succeq\!\matzero.
\end{array}
\right.\!\!\!\!\IEEEeqnarraynumspace
\end{IEEEeqnarray}
This problem can be solved efficiently using standard methods from convex optimization~\cite{BV04}.
%
If the solution matrix $\bB_\opR^\SDR$ has rank one, then (SDR-QP) finds the exact solution to the problem (\OBQP) in~\fref{eq:problem_tweak_real}. If, however, the rank exceeds one, we have to extract a precoding vector $\vecx^\SDR$ that belongs to the discrete set~$\setX^B$.
As commonly done, one can obtain such vector by first performing an eigenvalue-decomposition of~$\matB_\opR^\SDR$ and by then quantizing the first $2B$ entries of the leading eigenvector~$\bmu_\opR$. To this end, let $\bmx_\opR^\SDR$ denote the real-valued counterpart of $\vecx^\SDR$, whose $b$th entry ($b = 1,\ldots,2B$) is given~by
\begin{align}
\lefto[\bmx_\opR^\SDR\right]_b = \sqrt{\frac{P}{2B}} \sign\lefto([\bmu_\opR]_{2B+1}\right) \sign\lefto([\bmu_\opR]_b\right). 
\end{align}
The multiplication by $\sign\lefto([\bmu_\opR]_{2B+1}\right)$ takes into account the potential sign change caused by~$\psi$. 
The $b$th entry of the resulting complex-valued precoded vector $\vecx^\SDR$ is obtained as follows:
\begin{IEEEeqnarray}{rCl} \label{eq:complex_to_real}
\lefto[\vecx^\SDR\right]_b = \lefto[\bmx_\opR^\SDR\right]_{b}+ j \lefto[\bmx_\opR^\SDR\right]_{B+b}
\end{IEEEeqnarray} 
for $b=1,\ldots,B$. We refer to this approach as SDR with a rank-one approximation (SDR1). 
Alternatively, we can obtain a precoding vector in $\setX^B$ using more sophisticated randomized procedures. See the survey article \cite{luo10a} for more details. We refer to this approach as SDR with randomization~(SDRr).

SDR enables the computation of approximate solutions to the NP-hard problem (QP) in polynomial time. Specifically, the worst-case complexity scales with $B^{4.5}$~\cite{luo10a}. 
However, SDR lifts the problem to a higher dimension: from~$2B$ dimensions to $(2B+1)^2$ dimensions. Furthermore, implementing the corresponding numerical solvers entails high hardware complexity~\cite{castaneda16a}. 
Recently, a hardware-friendly \emph{approximate} SDR solver for problems of dimension up to $B=16$ was proposed in \cite{castaneda16a}. However, the complexity of this solver still prevents its use for massive MU-MIMO systems with hundreds of antennas. Hence, we conclude that SDR is a suitable technique only for small to moderately-sized systems (e.g.,~16~BS antennas or less). For larger antenna arrays, alternative methods are necessary. One such method is described next.

\subsection{Squared $\ell_\infty$-Norm Relaxation}

We next present a novel method to approximately solving~\eqref{eq:problem_complex_1bit}, which avoids lifting the problem to a higher dimension and requires low complexity. 
We start by rewriting the real-valued optimization problem~\eqref{eq:problem_tweak_real}~as 
\begin{align} \label{eq:ett_problem}
\begin{array}{ll}
\underset{\vecb_\opR \in \opR^{2B}}{\text{minimize}} & \vecnorm{\vecs_\opR - \matH_\opR \vecb_\opR}_2^2 + \dfrac{2B U N_0}{P} \vecnorm{\vecb_\opR}^2_\infty \\
\text{subject to} & [\bmb_\opR]_1^2 = [\bmb_\opR]_b^2, \,  b = 2,\,,\dots,\,2B
\end{array}
\end{align}
where we used that $\vecnorm{\vecb_\opR}_2^2 = 2B\vecnorm{\vecb_\opR}_{\infty}^2$ under the constraint that~$[\bmb_\opR]_1^2 = [\bmb_\opR]_b^2$ for $b = 2,\,\dots,\,2B$. By dropping the nonconvex constraints $[\bmb_\opR]_1^2 = [\bmb_\opR]_b^2$ for $b = 2,\,\dots,\,2B$, we obtain the following convex relaxation of \eqref{eq:ett_problem}:
\begin{IEEEeqnarray}{rCl} \label{eq:linf2problem}
&&(\LINF) \quad \underset{\vecb_\opR\in \opR^{2B}}{\text{minimize}} \ \vecnorm{\vecs_\opR - \matH_\opR \vecb_\opR}_2^2 + \frac{2 B U N_0}{P} \vecnorm{\vecb_\opR}_{\infty}^2 \IEEEeqnarraynumspace
\end{IEEEeqnarray}
which, as we shall see, can be solved efficiently. To extract a feasible precoding vector $\vecx^\LINF \in \setX^B$ from the solution $\vecb_\opR^{\LINF}$ to the problem~\fref{eq:linf2problem}, we quantize the entries of the vector to the quaternary set~$\setX$ by computing
\begin{align}
\lefto[\vecx_\opR^{\LINF}\right]_b = \sqrt{\frac{P}{2B}}\sign\lefto(\lefto[\vecb_\opR^{\LINF}\right]_b\right)
\end{align}
for $b = 1, \ldots, 2B$, where $\vecx_\opR^{\LINF}$ is the real-valued counterpart of $\vecx^\LINF$.
As in~\eqref{eq:complex_to_real}, we then obtain the complex-valued precoded vector as follows:
\begin{IEEEeqnarray}{rCl}
\lefto[\vecx^\LINF\right]_b = \lefto[\bmx_\opR^\LINF\right]_{b}+ j \lefto[\bmx_\opR^\LINF\right]_{B+b}
\end{IEEEeqnarray}
for $b = 1, \ldots, B$.
There exist several numerical optimization methods that are capable of solving problems of the form of~($\LINF$) in \eqref{eq:linf2problem} in a computationally efficient manner. The most prominent methods are forward-backward splitting (FBS) \cite{parikh13a,goldstein16a} and Douglas-Rachford~(DR) splitting \cite{lions79a,eckstein92a}. In what follows, we develop a DR splitting method, which we refer to as squared-infinity norm Douglas-Rachford splitting~(SQUID). We define the two convex functions $g(\bmb_\opR)=\vecnorm{\vecs_\opR - \matH_\opR \vecb_\opR}_2^2$ and $f(\bmb_\opR)= \frac{2 B U N_0}{P} \vecnorm{\vecb_\opR}_{\infty}^2$, and~solve
\begin{align}
\underset{\vecb_\opR\in \opR^{2B}}{\text{minimize}} \ &  g(\bmb_\opR) +f(\bmb_\opR).
\end{align}
Let
\begin{align}
 \text{prox}_h (\bmw) & = \argmin_{\bmb_\opR\in\reals^{2B}} h(\bmb_\opR) + 
\textstyle \frac{1}{2} \|\bmb_\opR-\bmw\|_2^2  \label{eq:firstproxydef} 
\end{align}
define the proximal operator for the function $h(\cdot)$~\cite{parikh13a}.
By initializing $\bmb^{(0)}_\opR=\bZero_{2B \times 1}$ and $\bmc^{(0)}_\opR=\bZero_{2B \times 1}$, SQUID performs the following iterative procedure for $t=1,2,\ldots$ until convergence or until a maximum number of iterations has been reached:
\begin{align}
\bma^{(t)}_\opR & = \text{prox}_g(2\bmb^{(t-1)}_\opR-\bmc^{(t-1)}_\opR) \label{eq:firstproxy}\\
\bmb^{(t)}_\opR & = \text{prox}_f(\bmc^{(t-1)}_\opR-\bma^{(t)}_\opR-\bmb^{(t-1)}_\opR)  \label{eq:secondproxy} \\
\bmc^{(t)}_\opR & = \bmc^{(t-1)}_\opR+\bma^{(t)}_\opR-\bmb^{(t-1)}_\opR.
\end{align} 
The proximal operator~$\text{prox}_g(\cdot)$ in~\fref{eq:firstproxy} has the following simple\footnote{One can further accelerate the evaluation of this proximal operator by using the Woodbury matrix identity (which reduces the dimension of the matrix inverse), and by precomputing certain constant quantities, such as $\bH^T_\opR\bms_\opR$.} expression:
\begin{align}
 \text{prox}_g (\bmw) = \textstyle (\bH^T_\opR\bH_\opR + \frac{1}{2}\bI_{2B\times2B})^{-1}(\bH^T_\opR\bms_\opR+\frac{1}{2}\bmw).
\end{align}
While the proximal operator for the $\ell_\infty$-norm is well known in the literature~\cite{parikh13a}, the proximal operator~$\text{prox}_f(\cdot)$ for the \emph{squared} $\ell_\infty$-norm, needed in~\fref{eq:secondproxy}, appears to be novel. The following theorem details an efficient procedure for computing this proximal operator. The proof is given in \fref{app:appC}.

\begin{thm} \label{thm:ellinfty2proximal}
Let  $\lambda>0$. Then, the squared $\ell_\infty$-norm proximal~operator 
\begin{align} \label{eq:proxyinfty2}
\bmu =  \text{prox}_{\lambda\ell_\infty^2} (\bmz)=  \argmin_{\bmu\in\opR^{N}} \lambda\|\bmu\|_\infty^2 + 
\textstyle \frac{1}{2} \|\bmz-\bmu\|_2^2
\end{align}
can be computed using the procedure summarized in~\fref{alg:proxinft2}.
\end{thm}

In summary, SQUID enables us to solve the relaxed problem in \fref{eq:linf2problem} in a computationally efficient manner. Indeed, each iteration requires only simple matrix and vector operations, and the evaluation of the proximal operator in \fref{alg:proxinft2}. The performance of SQUID is investigated in \fref{sec:simulation} where we demonstrate that this low-complexity algorithm achieves performance comparable to SDR, which is a far more demanding algorithm in terms of computational complexity.

\begin{algorithm}[t]
\caption{Proximal operator for the $\ellinf^2$-norm}\label{alg:proxinft2}
\begin{algorithmic}[1]
\State \textbf{inputs}: {$\vecz \in \opR^N,$ $\lambda \in (0,\infty)$}
\State $ \veca \gets \mathrm{abs}(\vecz)$ 
\State $ \vecs \gets \mathrm{sort}(\veca, $`descending'$)$  
\For{$k = 1,\ldots,N$}
\State $c_k \gets \frac{1}{2\lambda+k}\sum_{i=1}^k s_i $
\EndFor
\State $\alpha \gets\max\big\{0, \max_k\{{c_k}\}\big\}$
\For{$k = 1,\ldots,N$}
\State $u_k \gets \min\{a_k,\alpha\}\sign(z_k)$
\EndFor
\State \textbf{return} $\vecu$
\end{algorithmic}
\end{algorithm}

\subsection{Sphere Precoding} \label{sec:sphere}

Sphere decoding (SD) is a common method to solve CVPs exactly but at lower average computational complexity than a na\"ive exhaustive search \cite{fincke85a,agrell02a,studer10a}. 
The idea of SD is to constrain the search for possible optimal solutions to a hypersphere of radius~$r$. By transforming the optimal CVP into a tree-search problem, one can then perform a depth-first branch-and-bound procedure and prune branches that exceed the radius constraint to reduce the number of candidate vectors. While SD reduces (often significantly) the average complexity compared to an exhaustive search, it was shown to exhibit exponential complexity in the number of variables for data detection in multi-antenna wireless systems \cite{jalden05a,seethaler11a}.

To adapt SD to 1-bit quantized precoding (we call this adaptation \emph{sphere precoding} (SP)), we proceed as follows. Assume that the optimal precoding factor $\beta$ is known. Then, we can rewrite the objective function in \eqref{eq:problem_complex_beta_optimal} as follows:
\begin{IEEEeqnarray}{rCl}
\vecnorm{\vecs - \beta \matH\vecx}^2_2 + \beta^2 U N_0
&{=}& \vecnorm{\vecs \!-\! \beta \matH\vecx}^2_2 + \beta^2 \frac{U N_0}{P} \vecnorm{\vecx}_2^2 \label{eq:banan1} \IEEEeqnarraynumspace\\
&{=}& \vecnorm{\bar{\vecs} \!-\! \beta \overline{\matH}\vecx}^2_2 \label{eq:banan2}.
\end{IEEEeqnarray}
In~\eqref{eq:banan1}, we used that $\vecnorm{\vecx}_2^2 = P$ in the 1-bit case; in~\eqref{eq:banan2} we set~$\bar{\vecs} = [\vecs^T \ \veczero_B^T]^T$ and~$\overline{\matH} = [\matH^T \ \sqrt{{U N_0}/{P}} \, \matI_B]^T$. Hence, we can write the precoding problem as
\begin{align}
\underset{\bmx \in \setX^B}{\text{minimize}} \ & \vecnorm{\bar{\vecs} -\beta\overline{\matH}\vecx}_2
\end{align}
which can be solved using SD.  More specifically, by computing the QR factorization $\overline{\bH}=\bQ\bR$, where $\bQ\in\complexset^{(U+B)\times B}$ with~$\bQ^H\bQ=\bI_{B}$ and $\bR\in\complexset^{B\times B}$ is upper triangular with non-negative diagonal entries, we obtain the equivalent problem
\begin{align}
(\SP) \quad \underset{\bmx \in \setX^B}{\text{minimize}} \ & \vecnorm{\bQ^H\bar{\vecs} -\beta\matR\vecx}_2.
\end{align}
The triangular structure of this problem allows us to deploy standard SD methods, as the one in \cite{agrell02a}.

In practice, the optimal precoding factor $\beta$ is unknown. We therefore propose the following alternating optimization approach. 
At iteration $t=1$, we initialize the algorithm with the precoding factor obtained from WF precoding. Specifically, we use \eqref{eq:beta_optimal} and set $\beta_{1} = \hat\beta(\vecx^\text{WF})$.
We then solve ($\SP$) to obtain $\bmx_t^\SP$ and compute an improved precoding factor $\beta_{t+1} = \hat\beta(\bmx_t^\SP)$ using \eqref{eq:beta_optimal}. We repeat this procedure for~$t=2,3,\ldots$ until convergence or until a maximum number of iterations is reached. 
Our simulations have shown that this procedure usually converges in only $1$ to $3$ iterations and achieves near-optimal performance for small to moderately-sized MIMO systems (in \fref{sec:errorrate}, we present numerical results for the case of~$B = 8$ antennas). We note that a plethora of SD-related methods can be used to solve $\SP$. However, the exponential complexity of SD prevents its use for massive MIMO systems with hundreds of antennas.

\subsection{Decoding at the UEs} \label{sec:uedecoding}

As for the case of linear-quantized precoders, we assume that the $u$th UE is able to scale the received signal by some scaling factor $\beta_u$.
Note that the scaling factor~$\beta_u$ can not directly be chosen to be equal to the precoding factor~$\beta$, since~$\beta$ depends in the nonlinear case on the instantaneous transmit vector~$\vecs$ and cannot be estimated at the UEs. It is worth noting that for the special case in which the entries of~$\vecs$ are taken from of a constant-modulus constellation (e.g.,~$M$-PSK) and the receiver employs symbol-wise nearest-neighbor decoding (i.e., each UE maps its estimate~$\hat{s}_u$ in~\eqref{eq:scaling} to the nearest constellation point, which implies that both the residual MUI and the quantization error are treated as Gaussian noise, although they are not Gaussian), the scaling factor $\beta_u$ chosen by the receiver does not affect performance because the decision regions are circular sectors in the complex plane.
In the simulation results in \fref{sec:simulation}, we shall focus on QPSK modulation for which no scaling is needed.
In a follow-up work~\cite{jacobsson16d}, we presented simulation results for the case of higher-order constellations that do not satisfy the constant-modulus assumption (e.g., 16-QAM). In this case, it is sufficient to modify the precoding problem \eqref{eq:problem_complex_1bit} so that a single value of $\beta$ is chosen for a block of transmit symbols whose length does not exceed the channel coherence time. This allows the UEs to estimate $\beta$ through pilot transmissions or blind estimation~techniques.


\section{Numerical Results}
\label{sec:simulation}

We now present numerical simulations for the quantized precoders introduced in \fref{sec:linearprecoders} and \fref{sec:nonlinearprecoders}. 
Due to  space constraints, we shall focus on a limited set of system parameters.\footnote{Our simulation framework is available for download from GitHub (\url{https://github.com/quantizedmassivemimo/1bit_precoding}). The purpose is to enable interested readers to perform their own simulations with different system parameters and also to test alternative algorithms.}

\subsection{Error-Rate Performance}
\label{sec:errorrate}

We start by comparing the performance of the developed precoders in terms of uncoded bit error rate~(BER). In what follows, we assume that the UEs perform symbol-wise nearest-neighbor decoding.

\begin{figure}[t]
\centering
\subfloat[$B=8$ and $U=2$.]{\includegraphics[width=.9\columnwidth]{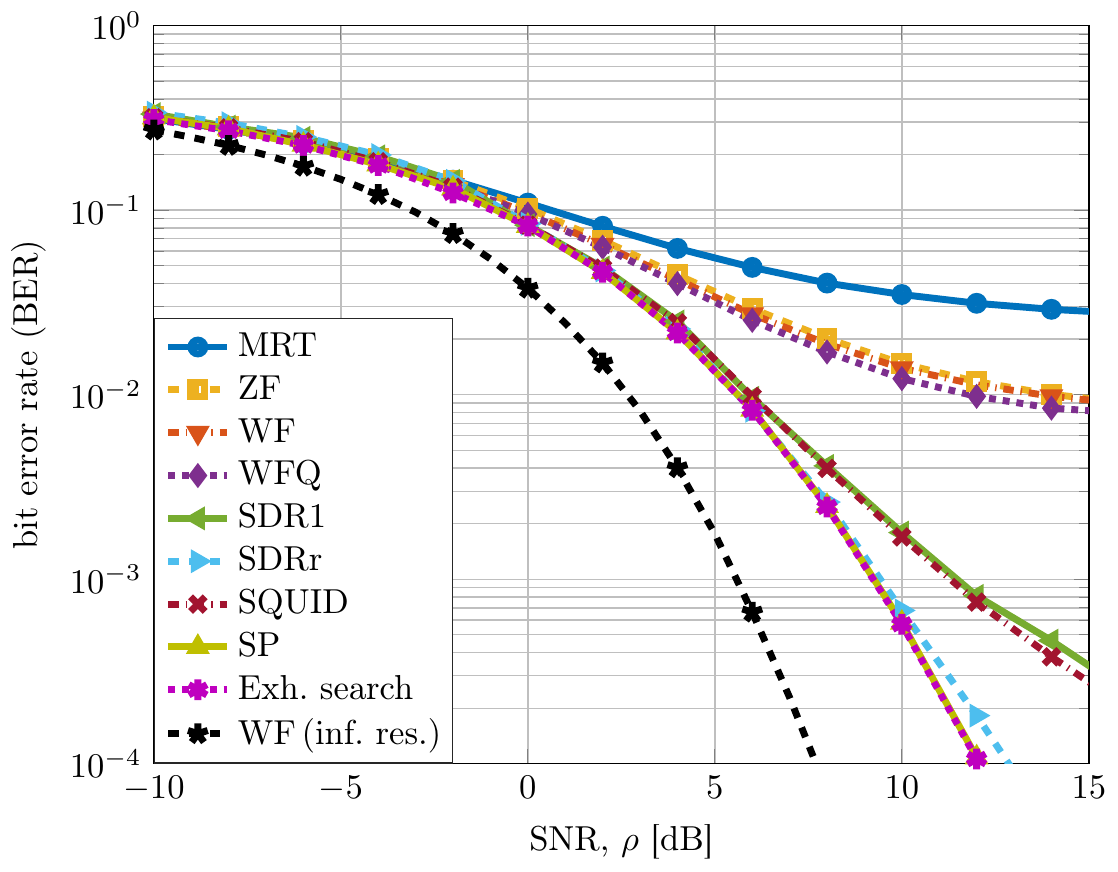}\label{fig:BER_ALL_QPSK_2x8}}
\quad
\subfloat[$B=128$ and $U=16$.]{\includegraphics[width=.9\columnwidth]{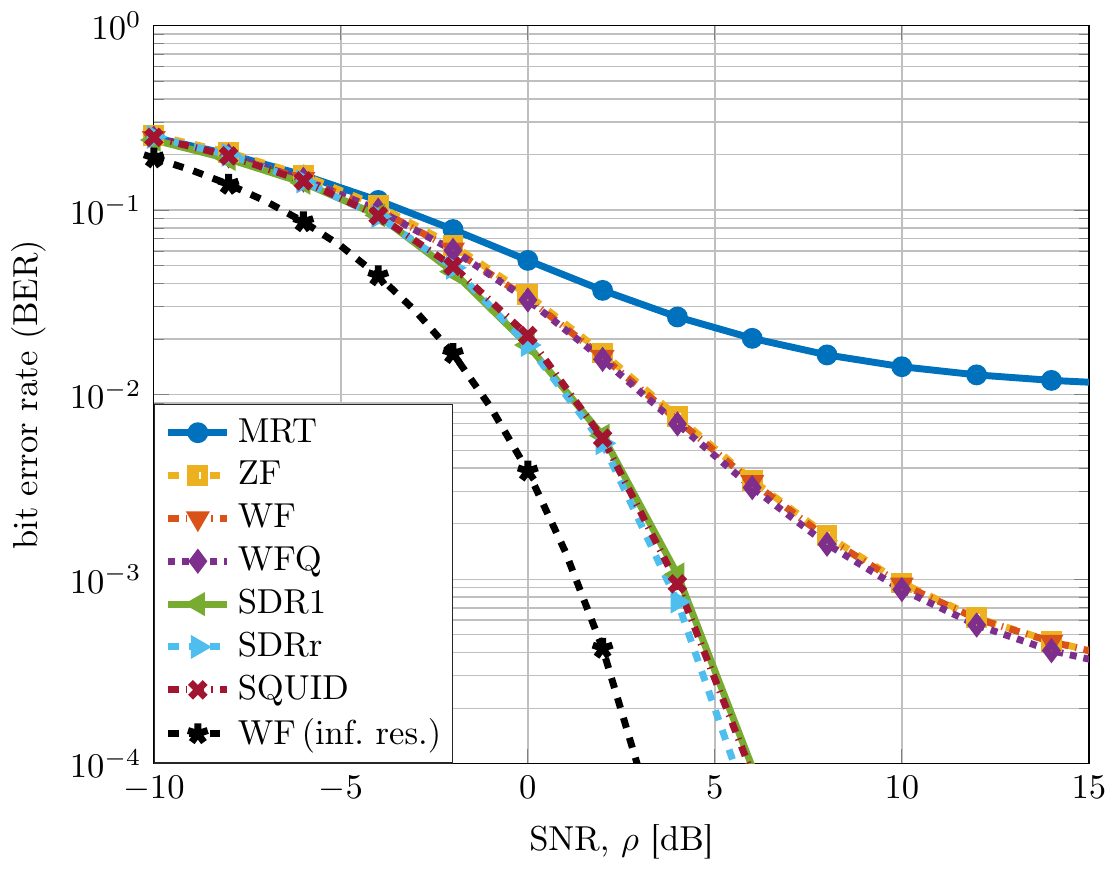}\label{fig:BER_ALL_QPSK_16x128}}
  \caption{Uncoded BER with QPSK signaling for 1-bit DACs as a function of the SNR, $\snr$, for the precoders introduced in~\fref{sec:linearprecoders} and in~\fref{sec:nonlinearprecoders}. The performance of the WFQ precoder proposed in~\cite{mezghani09c} is also illustrated for comparison.}
  \label{fig:BER_ALL_QPSK}
\end{figure}


In \fref{fig:BER_ALL_QPSK}, we compare the BER with QPSK signaling and 1-bit DACs for the linear precoders presented in~\fref{sec:linearprecoders} (namely, WF, ZF and, MRT) and the nonlinear precoding algorithms presented in \fref{sec:nonlinearprecoders} (namely, SDR1, SDRr, SQUID and SP). For comparison, we also report the performance of the WF-quantized (WFQ) precoder proposed in~\cite{mezghani09c}, and the performance of the WF precoder for the infinite-resolution~case.

In \fref{fig:BER_ALL_QPSK_2x8}, we consider the case $B=8$~BS antennas and $U=2$~UEs (moderately-sized MIMO system). For this case, one can find the optimal solution to (\OBQP) in \eqref{eq:problem_complex_1bit} by exhaustive search. 
We find that the gap between the performance of the optimal nonlinear precoder and the performance of the infinite-resolution WF precoder is remarkably small: about $4$~dB for a target BER of $10^{-3}$. 
Furthermore, the SP algorithm achieves near-optimal performance, as does SDRr. SQUID and SDR1 follow closely the optimal curve up to a BER of $10^{-2}$ but then their performance degrades.
The linear-quantized precoders, on the other hand, are adversely impacted by the coarse 1-bit quantization. Indeed, the BER for linear-quantized precoding saturates at~$10^{-2}$ or above.
Hence, in contrast to recently reported findings~\cite{saxena16a}, our results suggest that nonlinear precoding offers significant advantages in terms of BER compared to linear-quantized precoding.

In~\fref{fig:BER_ALL_QPSK_16x128}, we consider a massive MIMO system with~$B=128$~BS antennas and~$U=16$~UEs. Exhaustive search and SP are not viable in this setup due to the exponential complexity in $B$ that these methods entail. 
We note that the increased number of antennas yields a performance improvement for the linear-quantized precoders. Indeed, with ZF, WF, or WFQ one can now support error probabilities below~$10^{-3}$. 
However, the nonlinear precoders still significantly outperform the linear-quantized precoders.
The gap to the infinite-resolution BER with SQUID is about $3$~dB for a target BER of $10^{-3}$. With WFQ precoding, the gap is about $8$~dB for the same BER~target.

It is worth pointing out that at low SNR, the error-rate performance of all precoders is comparable. In this regime, linear-quantized precoders may offer a better performance-complexity trade-off. Furthermore, linear-quantized precoders may yield satisfactory BER performance for a larger range of SNR values if the number of BS antennas is increased further. 


\begin{figure}[t]
\centering
\includegraphics[width=.9\columnwidth]{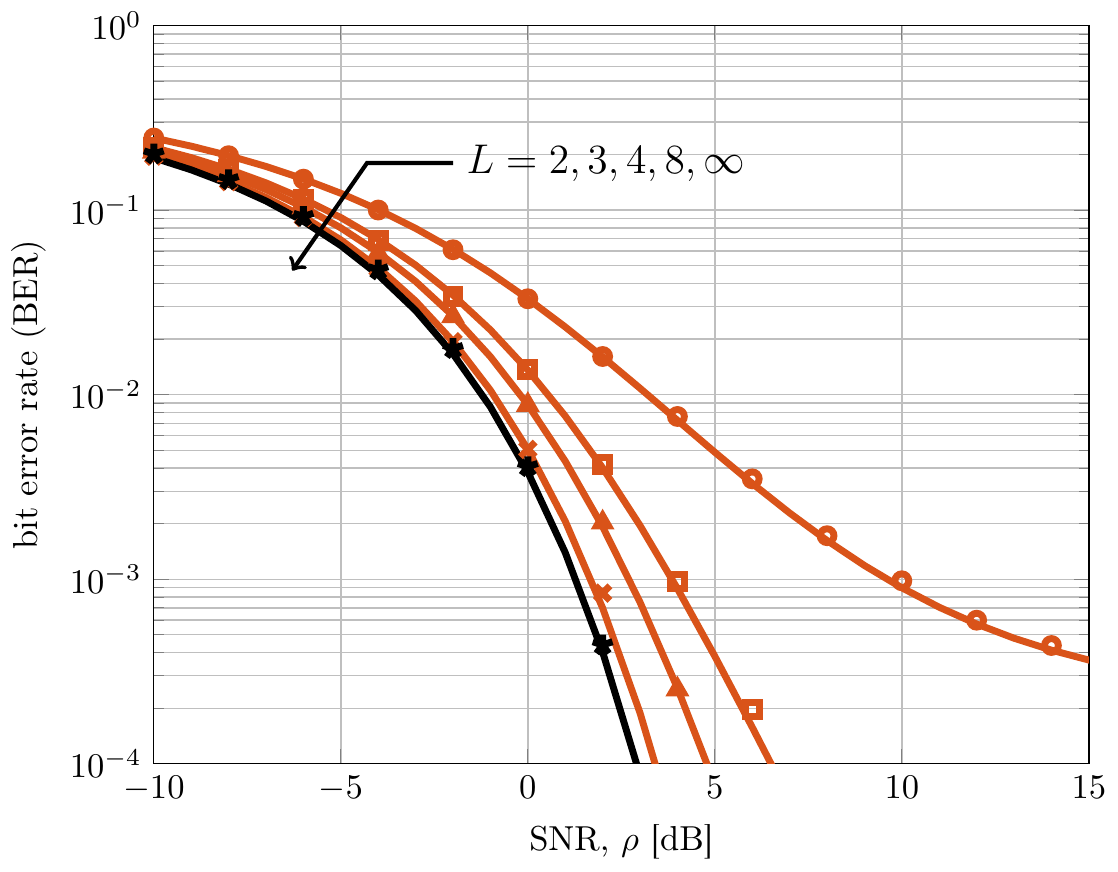}
\caption{Uncoded BER with QPSK signaling and WF precoding with multi-bit DACs;~$B = 128$ and~$U = 16$. Here, $L$ denotes the number of quantization levels. The markers correspond to simulated values and the solid lines correspond to closed-form approximations.}
\label{fig:BER_LMMSE_QPSK_16x128}
\end{figure}

In \fref{fig:BER_LMMSE_QPSK_16x128}, we show the uncoded BER for WF precoding as a function of the SNR and the number of DAC levels $L$ for a system with $U = 16$~UEs and $B = 128$~BS antennas. 
The simulated BER values in \fref{fig:BER_LMMSE_QPSK_16x128} are compared with closed-form approximations obtained by approximating the uncoded BER by~$1 - \Phi\lefto(\sqrt{\bar\sindr^\text{WF}}\right)$ where $\bar\sindr^\text{WF}$ is given in \eqref{eq:SINDR_WF}.
We observe that this approximation is accurate for the entire range of SNR values.
We further observe that low BER probabilities can be attained with very coarse DACs. Interestingly, by only adding a zero-level in the DACs (so that $L = 3$), the performance is drastically improved compared to the $1$-bit case ($L=2$). Furthermore, with only $3$-bit DACs ($L=8$) the performance gap to the infinite-resolution case is negligible. This suggests that it is possible to significantly reduce the number of bits in the high-resolution DACs used in today's systems.

\subsection{Robustness to Channel-Estimation Errors}
\label{sec:imperfectCSI}

So far, we have assumed that the BS has access to perfect CSI. In this section we shall relax this assumption to investigate the robustness of the developed algorithms to channel estimation errors. More specifically, we shall assume that the BS has access to a noisy version of $\matH$ modelled as
\begin{IEEEeqnarray}{rCl}
\widehat{\matH}=\sqrt{1-\varepsilon}\matH + \sqrt{\varepsilon}\matZ.
\end{IEEEeqnarray}
Here, $\varepsilon\in[0,1]$ and $\matZ$ has $\jpg(0,1)$ entries. We refer to $\varepsilon$ as the channel-estimation error:~$\varepsilon = 0$ corresponds to perfect CSI and $\varepsilon = 1$ corresponds to no CSI; intermediate values corresponds to partial~CSI.

\begin{figure}[t]
\centering
\includegraphics[width=.9\columnwidth]{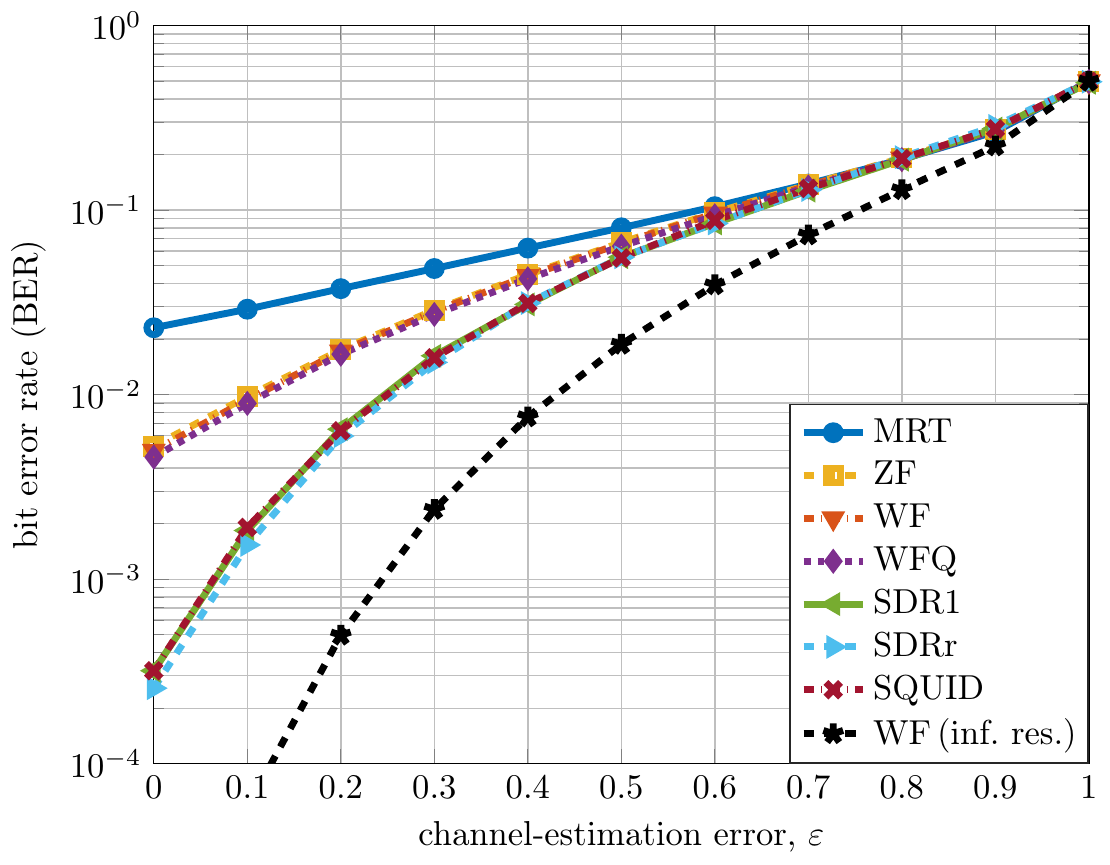}
\caption{Uncoded BER with QPSK signaling for 1-bit DACs as a function of the channel-estimation error,~$\varepsilon$.}
\label{fig:ESTERR_ALL_QPSK_16x128}
\end{figure}

In \fref{fig:ESTERR_ALL_QPSK_16x128}, we show, for the $1$-bit case, the uncoded BER with QPSK signaling as a function of the channel-estimation error $\varepsilon$ for a system with~$B = 128$~BS antennas and~$U = 16$~UEs.
Interestingly, the nonlinear precoders outperform the linear-quantized precoders whenever $\varepsilon \le 0.5$. This implies that nonlinear precoders can be used also when only imperfect CSI is available to the BS.

\subsection{Achievable rate} \label{sec:simrate}

Next, we validate the analytic results on the achievable rate with linear-quantized precoders reported in \fref{sec:linearprecoders} by numerical simulations.

\begin{figure}[t]
\centering
\includegraphics[width=.9\columnwidth]{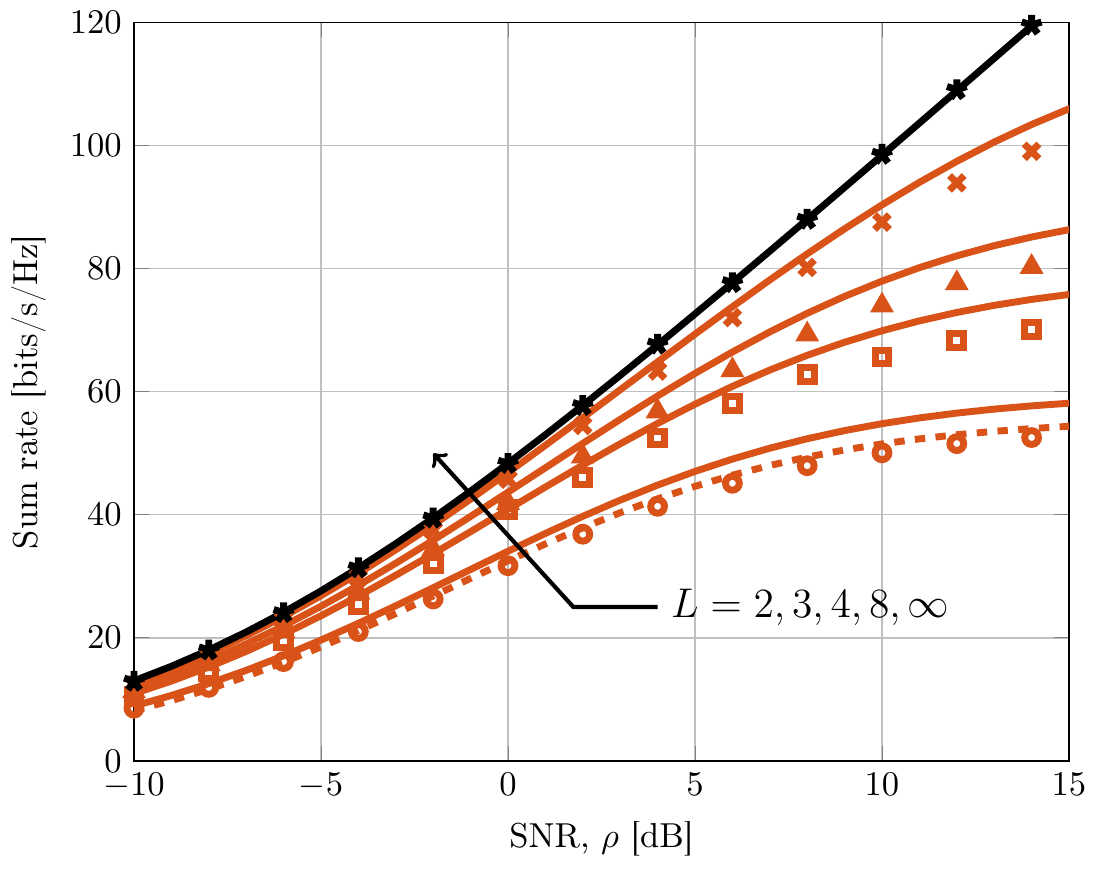}
\caption{Achievable sum-rate with Gaussian signaling and WF precoding with multi-bit DACs;~$B = 128$ and~$U = 16$. Here,~$L$ denotes the number of quantization levels. The markers correspond to simulated values and the solid lines correspond to closed-form approximations. The dashed line corresponds to the lower bound~\eqref{eq:rate_lower_bound} for 1-bit DACs.}
\label{fig:RATE_LMMSE_16x128_multibit}
\end{figure}

In \fref{fig:RATE_LMMSE_16x128_multibit}, we show the achievable sum-rate with Gaussian signaling and WF precoding as a function of the SNR and the number of DAC levels.
The rate approximation computed using \eqref{eq:SINDR_WF} is illustrated together with the rate lower bound \eqref{eq:rate_lower_bound} for the 1-bit case. 
We also show the achievable rate computed numerically using \eqref{eq:Ru_def} by simulating many noise and interference realizations for each channel realization and by mapping the resulting $\hat{s}_u$ to a rectangular grid in the complex plane to estimate the probability density functions required to compute~\eqref{eq:Ru_def} (see, e.g., \cite{jacobsson17b} for details).
We note that the asymptotic approximation matches well the numerical results, confirming its accuracy. We further note that, analogously to the uplink scenario \cite{jacobsson17b, li17b}, high sum-rate throughputs can be achieved despite having low-resolution DACs at the BS.

\section{Conclusions}
\label{sec:conclusions}
We have presented novel algorithms for the problem of downlink precoding in massive MIMO systems equipped with low-resolution DACs at the BS. To handle the challenges imposed by the finite-cardinality outputs of the DACs, we have considered two distinct approaches, namely linear-quantized precoding and nonlinear precoding.
%
%
We have shown that, with linear-quantized precoding, the use of DACs with $3$~to~$4$~bits of resolution is sufficient to close the performance gap (measured in terms of both BER and achievable rate) to the infinite-resolution case.
Furthermore, we have developed an asymptotic approximation of the effective SINDR, which can be used to predict the system performance accurately using simple closed-form expressions.

Linear-quantized precoders are, however, far from optimal. For the case of 1-bit DACs, we have shown that the error-rate performance can be significantly improved by allowing for nonlinear precoding. For example, we showed that for a BS with $128$~BS antennas serving $16$~UEs, the gap to infinite-resolution performance is about $8$~dB for a target $\text{BER}$ of $10^{-3}$ with linear-quantized precoders, but only $3$ dB with nonlinear precoders.
Nonlinear precoding, however, entails increased signal-processing complexity.
For small-to-moderate sized systems (e.g., 16~BS antennas or less), SDR- and SP-based precoders offer near-optimal BER-performance at tolerable complexity.
For massive MIMO systems, SQUID is an efficient and hardware-friendly algorithm to find a near-optimal solution to the 1-bit quantized precoding problem. 
In a follow-up study~\cite{castaneda17a}, we recently proposed two additional precoding algorithms for massive MIMO, and provide both very-large-scale integration (VLSI) designs and field-programmable gate array (FPGA) implementations. These designs demonstrate that nonlinear precoding algorithms can be realized in practice, with a manageable implementation complexity.

There are many avenues for future work. Extending our analysis and our algorithms to the frequency-selective case, where the use of orthogonal-frequency division multiplexing~(OFDM) is assumed, is part of ongoing work. {Early results for the case of OFDM and linear-quantized precoding are reported in~\cite{jacobsson17c}.}
As mentioned in \fref{sec:introduction}, the use of low-resolution DACs operating at symbol rate may result in significant out-of-band emissions and intersymbol interference. A characterization of both effects is critically required to assess the full potential of low-resolution DAC architectures in real-world MU-MIMO systems. Also, a generalization of our analysis to the case of oversampled DACs, which operate at a sampling frequency larger than the symbol rate, is of practical interest.


%


\appendices

%

\section{Proof of \fref{thm:decomp_uniform}}
\label{app:appB}

Let $\vecz = \matP\vecs \in \opC^B$ and $\vecx = \quantize(\vecz) \in \setX^B$. It follows from \fref{thm:bussgang} that the covariance matrices $\matC_{\vecx\vecz} = \Ex{\vecs}{\vecx\vecz^H}$ and~$\matC_{\vecz\vecz} = \Ex{\vecs}{\vecz\vecz^H}$ are related as follows:
\begin{IEEEeqnarray}{rCl} \label{eq:gain_matrix_bussgang}
\matC_{\vecx\vecz} &=& \matG \matC_{\vecz\vecz}
\end{IEEEeqnarray}
where $\matG$ is a $B \times B$ diagonal matrix with
\begin{IEEEeqnarray}{rCl} \label{eq:gain_antenna_gettingthere}
[\matG]_{b,b} &=& \frac{1}{\sigma_{b}^2} \Ex{}{\quantize(z_b)z_b^*}
\end{IEEEeqnarray}
where $z_b = [\vecz]_b$ and $\sigma_{b}^2 = \Ex{}{\lvert z_b \rvert^2}$  for $b=1,\ldots,B$. Note now that
\begin{IEEEeqnarray}{rCl}
\matC_{\vecz\vecz} &=& \Ex{\vecs}{\vecz\vecz^H} =\matP \Ex{\vecs}{\vecs\vecs^H} \matP^H = \matP\matP^H.
\end{IEEEeqnarray}
It follows from \eqref{eq:gain_matrix_bussgang} that we can write the quantized signal as $\vecx = \matG\vecz + \vecd$, where the distortion $\vecd$ is uncorrelated with $\vecz$. Indeed, note that
\begin{IEEEeqnarray}{rCl} \label{eq:uncorr_dist}
\Ex{\vecs}{\vecd\vecz^H} 
&=& \Ex{\vecs}{(\vecx \!-\! \matG\vecz)\vecz^H}
\!= \matC_{\vecx\vecz} - \matG \matC_{\vecz\vecz}
= \matzero_{B \times B} \IEEEeqnarraynumspace
\end{IEEEeqnarray}
where the last equality follows from~\eqref{eq:gain_matrix_bussgang}. 
We next evaluate~\eqref{eq:gain_antenna_gettingthere}. Note that, since the real and imaginary components of the symbol vector $\vecs$ are independent and identically distributed, so are the real and imaginary components of the precoded vector $\vecz$. Therefore, it holds that 
\begin{IEEEeqnarray}{rCl} \label{eq:Eqz}
\Ex{}{\quantize(z_b)z_b^*} = 2\Ex{}{\quantize(z)z}
\end{IEEEeqnarray}
 where we have introduced the random variable $z \sim \normal(0, \sigma_{b}^2/2)$. For a uniform DAC, the quantizer-mapping function can be expressed as
\begin{IEEEeqnarray}{rCl} \label{eq:quantizer_indicator}
\quantize(z) &=& \frac{\alpha\Delta}{2}(1-L) + \alpha\Delta \sum_{i = 1}^{L-1} \mathds{1}_{\lefto[\Delta\lefto(i - \frac{L}{2}\right), \infty\right)}(z).
\end{IEEEeqnarray}
Inserting~\eqref{eq:Eqz} and~\eqref{eq:quantizer_indicator} into~\eqref{eq:gain_antenna_gettingthere}, we get that
\begin{IEEEeqnarray}{rCl} 
[\matG]_{b,b}
&=& \frac{2}{\sigma_b^2} \Ex{}{\quantize(z)z} \\
&=& \frac{\alpha\Delta}{\sigma_b^2}(1-L)\Ex{}{z} \nonumber\\ &&+ \frac{2\alpha\Delta}{\sigma_b^2} \sum_{i = 1}^{L-1} \Ex{}{\mathds{1}_{\lefto[\Delta\lefto(i - \frac{L}{2}\right), \infty\right)}(z) z } \\
&=& \frac{2\alpha\Delta}{\sigma_b^2} \sum_{i = 1}^{L-1} \int_{\Delta\lefto(i - \frac{L}{2}\right)}^{\infty} \frac{z}{\sqrt{\pi\sigma_b^2}} \exp\lefto( -\frac{z^2}{\sigma_b^2}\right) \!dz \IEEEeqnarraynumspace\\
&=& \frac{\alpha\Delta}{\sqrt{\pi\sigma_b^2}} \sum_{i = 1}^{L-1} \exp\lefto( -\frac{\Delta^2}{\sigma_b^2} \lefto( i - \frac{L}{2}\right)^2 \right). \label{eq:gain_antenna_uniform}
\end{IEEEeqnarray}
Finally, the desired result~\eqref{eq:gainmatrix_uniform} follows from~\eqref{eq:gain_antenna_uniform} by using that $\sigma_b^2 = [\matP\matP^H]_{b,b}$.

\section{Proof of \fref{thm:ellinfty2proximal}}
\label{app:appC}

We start by rewriting the proximal operator in \fref{eq:proxyinfty2} as
\begin{IEEEeqnarray}{rCl} \label{eq:proxyinfty2new}
\bmu &=&  \argmin_{\bmx\in\reals^{N},~\alpha\in\reals} \lambda\alpha^2 + \frac{1}{2} \|\bmx-\bmy\|_2^2 \\ 
&&\text{subject to } x_k^2\leq\alpha^2, \quad k = 1,\ldots,N \nonumber
\end{IEEEeqnarray}
and use the Karush-Kuhn-Tucker (KKT) conditions \cite{BV04} to compute its solution. 
The Lagrangian of the optimization problem in \fref{eq:proxyinfty2new} is given by
\begin{IEEEeqnarray}{rCl}
\setL(\bmx,\alpha,\bmu) &=& \lambda\alpha^2 + \frac{1}{2}\|\bmx-\bmy\|^2_2 + \sum_{k=1}^N u_k(x_i^2-\alpha^2) \IEEEeqnarraynumspace
\end{IEEEeqnarray}
which yields the following two stationarity conditions:
\begin{align}
\lambda - \sum_{k=1}^N u_i  &= 0 \label{eq:stationarity1} \\
x_k-y_k + 2x_ku_k & =0, \quad k = 1,\ldots,N. \label{eq:stationarity2}
\end{align}
The stationarity condition \fref{eq:stationarity2} reveals that
$x_k = {y_k}/{(1+2u_k)}$,
which implies that if $u_k=0$, then $x_k=y_k$. Complementary slackness yields
$u_k(x_k^2-\alpha_k^2) = 0$,
which implies that if $u_k\neq0$, then $x_k^2=\alpha^2$ for a given~$k$. Hence, the values of $x_k$ must either be $|x_k|=\alpha$ or $x_k=y_k$ so that  $|x_k|<\alpha$. In words, the  proximal operator in \fref{eq:proxyinfty2new} clips to $x_k=\sign(y_k)\alpha$ the values $y_k$ whose magnitude exceeds $\alpha$ and leaves the remaining values unaffected. 
Hence, we only need to determine the optimal clipping threshold~$\alpha^*>0$. 

Assume $x_k\neq0$ without loss of generality (in the case $y_k=0$, we have $x_k=0$). Then, the stationarity condition in \fref{eq:stationarity2} reveals that
$u_k = \frac{1}{2}\!\left(\frac{y_k}{x_k}-1\right)\!.$
Together with the stationarity condition \fref{eq:stationarity1}, we have
\begin{align}
\sum_{k=1}^N u_k = \frac{1}{2}\sum_{k=1}^N \left(\frac{y_k}{x_k}-1\right) = \lambda
\end{align}
which implies that
\begin{align} \label{eq:importantfact1}
\sum_{k=1}^N \frac{y_k}{x_k} = 2 \lambda + N.
\end{align}
We now partition the indices $k=1,\ldots,N$ into two disjoint sets $\Omega$ and $\Omega^c$, The set $\Omega$ contains the indices of the entries $y_k$ for which $|y_k|\geq\alpha$; the set $\Omega^c$ contains the indices of the entries $u_k$ for which $|y_k|<\alpha$. 
Since~$x_k=\sign(y_k)\alpha$ for $k\in\Omega$ and $x_k=y_k$ for $k\in\Omega^c$, it follows from \fref{eq:importantfact1} that 
\begin{align}
\sum_{k\in\Omega} \frac{|y_k|}{\alpha} + \sum_{k\in\Omega^c} 1 = 2\lambda + N.
\end{align}
Hence,
\begin{align} \label{eq:abcde}
\sum_{k\in\Omega} \frac{|y_k|}{\alpha} = 2\lambda + N - |\Omega^c| = 2\lambda + |\Omega|.
\end{align}
We see from~\eqref{eq:abcde} that the clipping threshold $\alpha$ must satisfy
\begin{align} \label{eq:findmax}
\alpha = \frac{\sum_{k\in\Omega}|y_k|}{2\lambda+|\Omega|}.
\end{align}
To solve~\fref{eq:proxyinfty2new}, it is convenient to sort the values $|y_k|$ in descending order. 
Specifically, let us denote these values by $r_1\geq r_2\geq \dots \geq r_N$. 
Then one computes $\alpha_\ell=\sum_{k=1}^\ell r_k/(2\lambda+\ell)$ for $\ell=1,2,\dots,N$ and chooses $\alpha^*$ as the only $\alpha_{\ell}$ that satisfies $r_{\ell+1}<\alpha_{\ell}\leq r_{\ell}$. 
Simple algebraic manipulations reveal that this is equivalent to setting $\alpha^*=\max_{\ell} \alpha_{\ell}$. 
We then use $\alpha^* $ to perform element-wise clipping.
\fref{alg:proxinft2} implements exactly this procedure in a computationally efficient~manner.

%


\bibliographystyle{IEEEtran} 

\begin{spacing}{.9} 
\bibliography{IEEEabrv,confs-jrnls,publishers,studer,svenbib}
\end{spacing}

\end{document}